\documentclass[twocolumn,amsmath,amssymb, nobibnotes, superscriptaddress, aps]{revtex4-1}
\usepackage{graphicx}
\usepackage{varioref}
\usepackage{hyperref}
\usepackage{cleveref}
\usepackage{float}
\usepackage{color}
\usepackage[normalem]{ulem}

\newcommand{\SCO}{Sr\textsubscript{14}Cu\textsubscript{24}O\textsubscript{41}}

\begin{document}
	\title{Theoretical study of the spin and charge dynamics of two-leg ladders \\ 
               as probed by resonant inelastic x-ray scattering}
	\author{Umesh Kumar}
	\affiliation{Department of Physics and Astronomy, The University of Tennessee, Knoxville, TN 37996, USA}
	\affiliation{Joint Institute for Advanced Materials, The University of Tennessee, Knoxville, TN 37996, USA} 
	\author{Alberto Nocera} 
	\affiliation{Department of Physics and Astronomy, The University of Tennessee, Knoxville, TN 37996, USA}
	\affiliation{Materials Science and Technology Division, Oak Ridge National Laboratory, Oak Ridge, Tennessee 37831, USA}
	\affiliation{Stewart Blusson Quantum Matter Institute, University of British Columbia, Vancouver, British Columbia V6T 1Z4, Canada} 
	\author{Elbio Dagotto} 
	\affiliation{Department of Physics and Astronomy, The University of Tennessee, Knoxville, TN 37996, USA}
	\affiliation{Materials Science and Technology Division, Oak Ridge National Laboratory, Oak Ridge, Tennessee 37831, USA}
	\author{Steven Johnston}
	\affiliation{Department of Physics and Astronomy, The University of Tennessee, Knoxville, TN 37996, USA}
	\affiliation{Joint Institute for Advanced Materials, The University of Tennessee, Knoxville, TN 37996, USA} 
        \email{sjohn145@utk.edu}	
	\date{\today}
	
\begin{abstract}
Resonant inelastic x-ray scattering (RIXS) has become an
important tool for studying elementary excitations in correlated materials. 
Here, we present a systematic theoretical 
investigation of the Cu L-edge RIXS spectra of undoped and
doped cuprate two-leg spin-ladders in both the non-spin-conserving
(NSC) and spin-conserving (SC) channels.  The spectra are rich and host many
exotic excitations. In the NSC-channel of the undoped case, we identify
one-triplon and bound \textit{triplet} two-triplon excitations in the strong-rung coupling limit,
as well as confined spinons in the weak-rung coupling limit. 
In the doped case, we observe a quasiparticle excitation formed from a 
bound charge and spin-$\frac{1}{2}$ in the strong-rung coupling limit. 
In the SC-channel, we also identify several new features, including
bound \textit{singlet} two-triplon excitations  and confined spinons in
the undoped ladders in the strong- and weak-rung coupling limits, respectively.
Conversely, in the doped case, the SC channel primarily probes both gapless and gapped charge excitations. 
Finally, we revisit the available data for the ladder compound Sr$_{14}$Cu$_{24}$O$_{41}$ in the context of  
our results.  
\end{abstract}    

\maketitle    
\section{\label{Intro}Introduction\protect}
Strongly correlated spin ladders are excellent platforms for studying
quantum many-body phenomena, such as high critical  temperature (high-T\textsubscript{c})
superconductivity~\cite{Dagotto1D2D} and spinon confinement~\cite{Lake2009}.
Quantum ladders are intermediate between one- and two-dimensional materials,
and their study allows for detailed comparisons between theoretical models and experimental
probes~\cite{Schlappa2012, FSDIR, PhysRevLett.103.047401, MartinNaturePhys, 
WhitePRL2003, PhysRevB.85.224436, PhysRevLett.108.167201, PhysRevLett.118.167206, 
PhysRevB.88.094411, PhysRevLett.102.107204, PhysRevLett.81.1702, PhysRevB.62.8903, 
PhysRevLett.105.097202, PhysRevB.88.014504, PhysRevB.80.094411, PhysRevB.83.140413, 
PhysRevB.89.174432, Lake2009, PhysRevB.97.195156}.
The discovery of superconductivity in the ladder ``telephone number''  compound
Sr$_{0.4}$Ca$_{13.6}$Cu$_{24}$O$_{41.84}$~\cite{doi:10.1143/JPSJ.65.2764},
which had been theoretically predicted~\cite{PhysRevB.45.5744}, 
created new opportunities to study the relationships between lattice,
orbital, charge, and magnetic degrees of freedom  and unconventional
superconductivity in copper-oxide materials. Accordingly, a significant effort has been launched to understand the magnetic excitation spectrum of materials hosting quantum ladders and its connection to superconductivity.  For example, inelastic neutron scattering (INS) studies have reported the
observation of a spin gap in 
Sr$_{14}$Cu$_{24}$O$_{41}$~\cite{PhysRevLett.81.1702}, triplon and two-triplon
excitations in La$_4$Sr$_{10}$Cu$_{24}$O$_{41}$~\cite{PhysRevLett.98.027403}, and spinon
confinement in CaCu$_2$O$_3$~\cite{Lake2009}.
 
With continued improvements in instrumentation, resonant inelastic x-ray
scattering (RIXS) is being increasingly employed to study collective magnetic 
excitations~\cite{PhysRevLett.104.077002, PhysRevB.85.214527,
PhysRevLett.103.047401,Schlappa2012,FSDIR,PhysRevB.88.020501,PhysRevLett.114.217003,Huang2016,LeTacon2011,Lee2014}.  
RIXS is complementary to INS in that 
the scattering processes allows for both $\Delta S = 0$ and 
$\Delta S = 1$ excitations, depending 
on the elemental edge~\cite{PhysRevB.85.214527,PhysRevLett.103.117003}, the 
strength of the spin-orbit coupling in the core level~\cite{PhysRevLett.103.117003}, and 
the local crystal stucture of the material~\cite{PhysRevB.97.041102}. 
As such, the technique accesses many magnetic excitations 
including magnons~\cite{LeTacon2011,Huang2016,PhysRevLett.114.217003,PhysRevB.88.020501,Lee2014}, 
bimagnons in two-dimensional (2D) cuprates~\cite{PhysRevB.85.214527,PhysRevB.85.214528}, 
and multi-spinon excitations in one-dimensional (1D) cuprates~\cite{PhysRevB.83.245133,PhysRevB.85.064423, 
FSDIR, KUMARNJP1}. RIXS also provided surprising results for 
2D cuprates, where the paramagnon excitations are found to persist deep
into the overdoped region of the phase diagram~\cite{LeTacon2011,Jia2014,PhysRevB.88.020501,PhysRevLett.114.217003,Huang2016}.  
Recently, an electron-hole asymmetry in the doping dependence of the spin 
excitations of 2D cuprates was reported, 
as well as an additional collective charge excitation in the electron-doped 
case that is absent in the hole-doped case~\cite{Lee2014}. 

The rich variety of excitations observed in 1D and 2D cuprates described above, 
and their possible connection to
unconventional superconductivity, provides a strong motivation for exploring the RIXS spectra of two-leg
spin ladders, both as a function of the rung coupling and doping.  
Such studies provide information not only about magnetic excitations but also about potentially cooperative/competing charge, orbital, and lattice excitations.  
Early RIXS Cu $K$-edge experiments on the telephone number compounds focused primarily on the high-energy 
charge excitations across the Mott gap~\cite{HigashiyaNJP,
PhysRevB.76.100507,PhysRevB.76.045124}. 
Later, as the instrumental resolution improved, studies started addressing low-energy magnetic excitations. 
For example, the magnetic response of Sr$_{14}$Cu$_{24}$O$_{41}$ at the Cu $L_3$-edge was 
measured~\cite{PhysRevLett.103.047401} and interpreted in terms of the lower boundaries of a two-triplon continuum. 
Another Cu $L_3$-edge study on CaCu\textsubscript{2}O\textsubscript{3} -- a weakly coupled spin-ladder system -- 
showed that the spectra could be decomposed into contributions from the 
spin-conserving (SC) and non-spin-conserving (NSC) channels~\cite{PhysRevLett.112.147401}. 
Subsequent work at the same edge on the same material focused 
on spin-orbital fractionalization, but did not carry out an analysis of 
spinon confinement~\cite{PhysRevLett.114.096402}.

From a theoretical perspective, studies of the RIXS response of spin ladders 
have mainly focused on undoped systems using a projector method~\cite{PhysRevB.85.224436} 
or exact diagonalization (ED) of small ($4\times 2$) Hubbard clusters~\cite{PhysRevLett.103.047401}, 
and were restricted to a limited set of rung couplings. 
To our knowledge, no systematic RIXS study of the low-energy excitations 
of doped and undoped ladders has been carried out. 
Here, we present such a study. Specifically, we use the Kramers-Heisenberg formalism to compute the  
RIXS response of undoped and doped two-leg $t$-$J$ ladders 
while varying the superexchange coupling along the
rungs over a wide range of values. The RIXS intensity is evaluated numerically, exactly or with a very small error, using 
ED and the density matrix renormalization group (DMRG)~\cite{NoceraSR1, PhysRevLett.69.2863, PhysRevB.48.10345} methods.  
Using these tools, we investigate the charge and magnetic excitations in both the
SC and NSC channels and catalog an assortment of quasiparticle and collective excitations. 
The present systematic study can guide future RIXS experiments and help to classify compounds 
as being in the weak- or strong-rung coupling regime, 
depending on the observed excitations.   

Iridates provide another group of spin-$\frac{1}{2}$ ($J_\text{eff}=\frac{1}{2}$) materials 
that have been studied with RIXS~\cite{PhysRevLett.101.076402, PhysRevB.84.020403, PhysRevLett.108.177003, PhysRevLett.117.107001}.
Moreover, progress was recently made in engineering quasi-1D iridates in heterostructures~\cite{doi:10.1002/adma.201603798}, establishing
another platform for examining and controlling the properties of quantum spin ladders. 
The results presented here can serve as a valuable roadmap in these contexts as well.      

Our organization is as follows: Section~\ref{methods} introduces the spin
ladder model and the relevant scattering cross-sections for RIXS within the Kramers-Heisenberg formalism. 
Sections~\ref{NSCchannel} and \ref{SCchannel} present results for the 
RIXS spectra in the NSC and SC channels, 
respectively. Section~\ref{Schlappadata} 
revisits and discusses the RIXS data reported~\cite{PhysRevLett.103.047401} on the spin-ladder
compound~ Sr$_{14}$Cu$_{24}$O$_{41}$ in the context of our results. 
Section~\ref{conclusion} summarizes our findings. 
	
\section{Methods\label{methods}}
\subsection{Model Hamiltonian} 
We study the $t$-$J$ model in a two-leg ladder geometry. The Hamiltonian is
\begin{equation}\label{tJmodel} \begin{split} H&= J_\text{rung}\sum_{i} \left(\textbf{S}_{i,0}\cdot \textbf{S}_{i,
1}-\tfrac{1}{4}n_{i,0} n_{i,1}\right) \\ &+J_\text{leg}\sum_{i,\tau}
\left(\textbf{S}_{i,\tau}\cdot \textbf{S}_{i+1,\tau}-\tfrac{1}{4}n_{i,\tau}
n_{i+1,\tau}\right)  \\ &+t_\text{rung}\sum_{i,\sigma}\left(c_{i,0,\sigma}^\dagger
c^{\phantom{\dagger}}_{i,1,\sigma}+\text{h.c.}\right)\\ &+t_\text{leg}\sum_{i, \tau,\sigma}\left(c_{i,\tau,\sigma}^\dagger
c^{\phantom{\dagger}}_{i+1,\tau,\sigma}+\text{h.c.}\right).\\ \end{split}
\end{equation} 
Here, $\tau = 0,~1$ indexes the legs of the ladder while 
$i=1,\dots,L$ indexes the unit cell along each leg; $\textbf{S}_{i,\tau}$ is a spin operator;
$c_{i,\tau,\sigma}^{\phantom{\dagger}}~(c_{i,\tau,\sigma}^\dagger)$
annihilates~(creates) a hole with spin $\sigma$ ($=\uparrow,\downarrow$) at site ($i, \tau$)
subject to the constraint of no double occupancy; $J_\text{
leg}~(J_\text{rung})$ and $t_\text{leg}~(t_\text{rung})$ are
the superexchange and hopping integrals along the leg~(rung) direction of the
ladder, respectively; 
and $n_{i,\tau}= \sum_\sigma c^\dagger_{i,\tau,\sigma}c^{\phantom\dagger}_{i,\tau,\sigma}$ is the 
hole number operator. Note that we neglected the ring-exchange terms in our model for simplicity.

The two-leg spin-ladder model can be used to describe several compounds, and a 
range of model parameters have been reported, as 
summarized in Table \ref{Tbl:Parameters}. The parameters can be different even for the same compound 
depending on the nature of the experiment or model used to analyze the data.
Due to the variability in the reported couplings, we opted to carry out a 
systematic study over a range of rung parameters spanning from \textit{weak} to 
\textit{strong} rung couplings. 
Unless otherwise stated, we adopt the specific couplings (in units of meV) $J_\text{leg}=140$,
 $J_\text{rung}=140r$, $t_\text{leg}=-300$, and $t_\text{rung}=-300\sqrt{r}$, where 
$r=\frac{J_\text{rung}}{J_\text{leg}}$ is a parameter 
used to adjust the ratio of the rung-leg couplings. The choice $t_\text{rung}=-300\sqrt{r}$ 
preserves the relationship $J_\text{rung}\propto \frac{t_\text{rung}^2}{U}$, assuming a 
fixed value for the Coulomb interaction $(U)$ strength in an effective Hubbard model. 

\begin{table}[t]
	\begin{tabular}{p{2.3cm}p{1.2cm}p{1.2cm}p{1.2cm}l}
		\hline
		\hline
		Material                          & $J_\text{leg}$ & $J_\text{rung}$                      & $J_\text{ring}$  & Ref.  \\ \hline
		Sr$_{14}$Cu$_{24}$O$_{41}$    & 130             & 72                                   &                  & INS~[\onlinecite{PhysRevLett.81.1702}].     \\ \hline
		& 110             & 140                                  &                  & RIXS~[\onlinecite{PhysRevLett.103.047401}]. \\ \hline
		& $110\pm 20$     & 4J$_\text{leg}$/5 &                  & Raman~[\onlinecite{PhysRevLett.87.197202}]. \\ \hline
		& $145$     & 123 &                  & This work. \\ \hline
		La$_4$Sr$_{10}$Cu$_{24}$O$_{41}$ & 186             & 124                                  & 31               & INS~[\onlinecite{PhysRevLett.98.027403}]. \\ \hline
		La$_6$Ca$_{8}$Cu$_{24}$O$_{41}$  & 110             & 110                                  & 16.5             & INS~[\onlinecite{PhysRevB.62.8903}].       \\ \hline
		CaCu$_{2}$O$_{3}$                & 134             & 11                                   &                  & RIXS~[\onlinecite{PhysRevLett.114.096402}]. \\ \hline\hline
	\end{tabular}
	\caption{\label{Tbl:Parameters} Different values of the exchange
        parameters (in units of meV) reported in the literature for various spin-$\frac{1}{2}$ ladder systems. 
        Entries where the value of $J_\text{ring}$ is missing 
        correspond to studies where the ring exchange terms were not included in the analysis.}
\end{table}	
 
\subsection{RIXS Intensity}
We evaluated the RIXS response at the Cu $L$-edge of cuprate materials. 
In a RIXS experiment, photons with energy $\omega_\text{in}$ and 
momentum $\textbf{k}_\text{in}$ ($\hbar=1$) scatter inelastically from a sample, 
transferring momentum $\textbf{q }= \textbf{k}_\text{out}-\textbf{k}_\text{in}$ and energy $\omega =
\omega_\text{out} -\omega_\text{in}$ to its elementary excitations. 
The RIXS spectrum is evaluated using the Kramers-Heisenberg
formula~\cite{RevModPhys.83.705} and is given by 
\begin{equation}
\mathcal{I} = \sum_{f}\bigg|\frac{\langle f|\mathcal{D}_{{\bf k}_\text{out}}^\dagger |n\rangle
\langle n|\mathcal{D}^{\phantom{\dagger}}_{\bf{k}_\text{in}}|g\rangle}{E_g+\omega_\text{in}-E_n+\mathrm{i}\Gamma}\bigg|^2\delta(E_f-E_g+\omega),  
\end{equation}
where $|g\rangle, |n\rangle$, and $|f\rangle$ are the ground, intermediate, and
final states with energies $E_g$, $~E_n$, and $E_f$, respectively, and $\Gamma$  is the core-hole lifetime broadening.
The eigenstates are obtained by diagonalizing $H + H_\text{ch}$, 
where $H_\text{ch} = V_c\sum_{i,\tau}  n^{\phantom{p}}_{i,\tau} n_{i,\tau}^p$ accounts for the 
interaction between the valence and the core holes in the intermediate state.  
Here, $V_c$ is the inter-orbital repulsion between the holes in the Cu $2p$ and $3d$ orbitals, 
$n_{i,\tau}^p=\sum_{\alpha}p_{i,\tau,\alpha}^\dagger p^{\phantom\dagger}_{i,\tau,\alpha}$, 
and $p_{i,\tau,\alpha}^\dagger~(p^{\phantom\dagger}_{i,\tau, \alpha})$ 
creates (annhilates) a hole in the $J=\tfrac{3}{2}~(\tfrac{1}{2})$ core level of site $(i,\tau)$ for $L_3~(L_2)$-edge. 
In the two-leg $t$-$J$ ladder, the dipole operator is given by $ \mathcal{D}_{\bf k}=\sum_{i,\tau,\sigma, \alpha}
e^{\mathrm{i}\textbf{k}\cdot\textbf{R}_{i,\tau}}[c^{\phantom{\dagger}}_{i,\tau,\sigma} p_{i, \tau, \alpha}^\dagger
+ \mathrm{h.c.}]$, where we have neglected the prefactor that depends on the polarization of the photon and the  
scattering angle. Due to the large spin-orbit coupling in the core 2$p$ orbital, both NSC ($\Delta S = 1$) and
SC ($\Delta S = 0$) excitations can occur in this edge,~\cite{PhysRevB.85.064423, NoceraSR1, PhysRevLett.112.147401} and 
the RIXS spectra has contributions from both  of these channels. However, 
it has been recently shown how the Cu $L_3$-edge spectra can be 
resolved into its individual SC and NSC components~\cite{PhysRevLett.112.147401}. For this reason, we will consider these two 
channels separately in what follows.   

The momentum transfer has two components in a two-leg ladder geometry: $\textbf{q}=(q_x, q_y)$, 
where $q_x= 2\pi n/La$, with $n\in[0,L)$ but $q_y = 0$ or $\pi/a$, only. 
For our ED calculations, we evaluate Eq.~(2) directly, while the 
details of our DMRG approach are given in Ref. \onlinecite{NoceraSR1}. 

Throughout this study, we use $V_c=6.7t$, $\Gamma = |t|$ 
for all $n$, and a Lorentzian broadening with $\eta = J_\text{leg}/6$ for the energy-conserving 
delta function appearing in Eq.~(2), unless stated otherwise. 
These parameters are typical for Cu $L$-edge measurements on the cuprates. Most of the spectra were computed using ED 
on $N = L\times 2 = 10\times 2$ clusters with periodic boundary conditions, while DMRG was used 
on $N = L\times 2 = 16\times 2$ undoped clusters with open boundary conditions. 
For the doped cases, our ED results are for a filling of $\langle n\rangle = 0.9$  
(or 10\% doping). Finally, the incident photon energy $\omega_\mathrm{in}$  was tuned to match 
the maximum of the x-ray absorption spectrum, as discussed in Appendix~\ref{XAScalculation}.  
	
\begin{figure}[t] 
	\centering    
	\includegraphics[width=\linewidth]{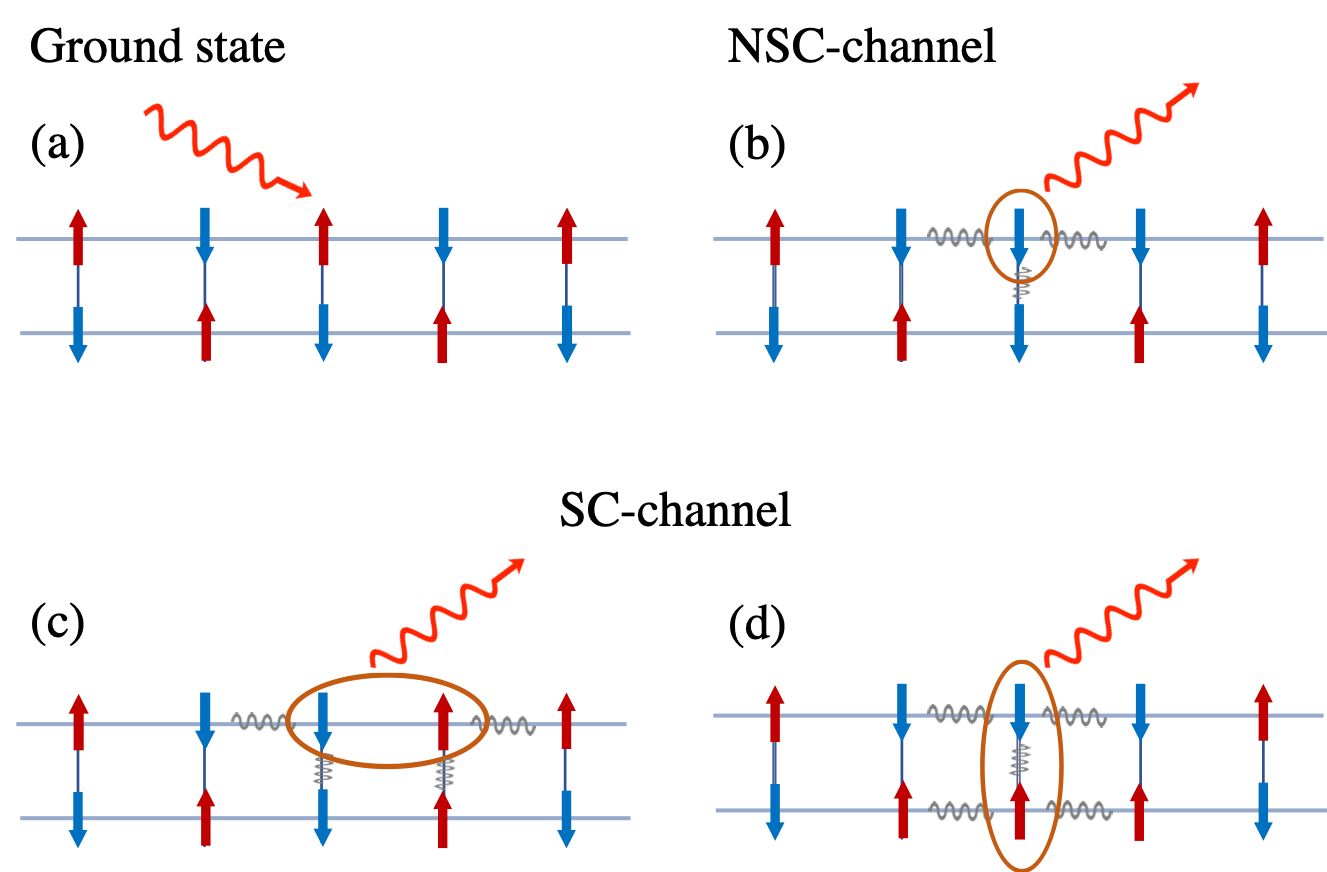}
	\caption{Schematic diagrams of the elementary magnetic excitations that are possible in spin-ladders. Panel (a)
        shows the ground state configuration of the spins with antiferromagnetic correlations. 
        Panel (b) shows the single spin-flip excitations 
        that appear in the non-spin-conserving (NSC) channel. Panels (c) and (d) show the
        double spin-flip processes relevant to the spin-conserving channel (SC). In each panel, the wiggly grey lines 
        indicate ``broken" magnetic bonds.   
        }     
	\label{fig:schematicladder}            
\end{figure}
	
\section{Results and Discussion \label{results}}

We calculated the RIXS spectra in various rung coupling regimes, 
ranging from \textit{strong} ($r=\frac{J_\text{rung}}{J_\text{leg}}=4$, $2$), to \textit{isotropic} ($r= 1$), 
to \textit{weak} ($r = 0.5, 0.25, 0.1$). But before examining our results, it is 
worthwhile to review the various excitations that are expected in a two-leg spin-ladder system. 

The magnetic excitations of undoped spin-$\frac{1}{2}$ ladders in the \textit{strong} rung
coupling limit are well understood starting from a dimerized rung
basis~\cite{PhysRevB.63.144410, PhysRevB.45.5744}. For $r \rightarrow \infty$, 
the individual rungs of the ladder are decoupled, each forming a spin dimer. 
For the antiferromagnetic case, the ground state of the $L$-rung ladder is then 
a direct product of rung singlets with total spin $S = 0$.  
The elementary excitations of this state are ``triplons''~\cite{PhysRevLett.90.227204,doi:10.1142/S0217984905009237,PhysRevB.47.3196},
where one or more of the rungs are excited into the triplet manifold~\footnote{The triplon and bound two-triplon 
excitations are often referred to as magnon and bound two-magnon excitations, respectively, in the literature. We have 
adopted the triplon nomenclature to be consistent with the previous RIXS studies~\cite{PhysRevLett.103.047401,
PhysRevB.85.224436}.}. 
For example, the first excited state is $L$-fold degenerate,  
where one of the rungs is in a triplet spin configuration, while the higher lying 
excited states involve integer numbers of rung triplets distributed throughout the system. 
The degeneracy of the excited states is lifted when $J_\text{leg}\ne0$, leading to a dispersive 
quasiparticle excitation. 
To order $\mathcal{O}\left(J^2_\text{leg}/J_\text{rung}\right)$, the triplon dispersion is~\cite{PhysRevB.47.3196}
\begin{equation}\label{Eq:Triplon}
\omega_t(q)= J_\text{rung}\Big[1+\frac{J_\text{leg}}{J_\text{rung}}\cos(qa) 
+ \frac{3}{4}\Big(\frac{J_\text{leg}}{J_\text{rung}}\Big)^2\Big],
\end{equation} 
where $a$ is the lattice constant along the leg direction. 

The two-triplon excitation manifold is even richer. 
Here, the two-triplon excitations appear in three angular momentum channels corresponding to~$S=0, 1, 2$, 
namely the \textit{singlet}, \textit{triplet} and \textit{quintet} channels,
respectively~\cite{refId0,PhysRevLett.81.1941, PhysRevB.63.144410}.   
Previous work~\cite{refId0} showed that a finite value of $J_\text{leg}$ 
can lead to two-triplon bound states whose dispersions 
in the singlet ($S = 0$) and triplet ($S=1$) channels are 
to order $\mathcal{O}\left(J^3_\text{leg}/J^2_\text{rung}\right)$ given by~\cite{PhysRevB.63.144410} 
\begin{equation} \label{striplon}
\begin{split}
&\omega_{2t}^S(q)= J_\text{rung}\Big[2-\tfrac{3}{2r}+\tfrac{19}{16r^2}  -\tfrac{9}{32r^3} -\big(\tfrac{1}{2r}- \tfrac{1}{8r^2}+\tfrac{51}{128r^3} \big) \\ 
& \times\cos(qa)-\big(\tfrac{5}{16r^2}+\tfrac{21}{32r^3}\big)\cos(2qa) -\tfrac{37}{128r^3}\cos(3qa)\Big],
\end{split}
\end{equation}
and
\begin{equation} \label{ttriplon}
\begin{split}
&\omega_{2t}^T(q) = J_\text{rung}\Big[2-\tfrac{3}{2r}+\tfrac{11}{8r^2}  +\tfrac{17}{16r^3} -\big(\tfrac{1}{r}+ \tfrac{1}{4r^2}-\tfrac{9}{16r^3} \big) \\
&\times\cos(qa)-\big(\tfrac{1}{2r^2}+\tfrac{1}{2r^3}\big)\cos(2qa) -\tfrac{5}{16r^3}\cos(3qa) \Big], 
\end{split}
\end{equation}
respectively. 

The two-triplon excitations in the \textit{triplet} channel were analyzed in a recent INS 
study~\cite{PhysRevLett.98.027403}, reporting a dispersive
excitation along the $\textbf{q}=(q_x,0)$ direction in the Brillouin zone, in
very good agreement with the \emph{lower boundary} line of the two-triplon
continuum. It was also argued that the four-spin ring-exchange term frustrates
the formation of a $S=1$ bound state below the continuum. 
In this effort, we neglect the four-spin cyclic
exchange, thus we find that our data overlays well with the dispersion of the
bound $S = 1$ two-triplon state. Our results are also consistent 
with the available RIXS experimental data~\cite{PhysRevLett.103.047401}. 
We know of no experimental study probing
two-triplon excitations in the \textit{singlet} and \textit{quintet} channels; however, 
our results below show that the RIXS SC channel can access
the two-triplon bound state in the \textit{singlet} channel. 
	
In the \textit{weak} rung coupling limit for undoped two-leg spin-$\frac{1}{2}$ ladders, the system 
can be viewed as a set of weakly coupled Heisenberg chains. In this regime, the  
excitation spectrum is understood in terms of a confined spinon continuum with a finite 
spin gap \cite{PhysRevLett.77.1865}. Intuitively, the excitations of a Heisenberg chain are spin-$\frac{1}{2}$ spinons, 
which always appear in pairs and are basically free to move along a single chain. 
When the two chains are coupled antiferromagnetically, however, 
the spinons within a single chain feel an effective confining potential~\cite{Lake2009}. 
This potential is created by the region of ferromagnetically coupled spins that forms on the rungs 
between the two spinons as they separate. 

When $r \ll 1$, the two-leg ladder problem can be mapped onto one of weakly interacting 
\textit{singlet} and \textit{triplet} Majorana fermions with effective masses 
$m_s = 3m$ and $m_t = m$, respectively, 
where $m\approx 0.41 J_\text{rung}$~\cite{PhysRevB.53.8521}. 
The excitation spectrum, as encoded in the dynamical spin structure factor 
$S(q,\omega)$~\cite{PhysRevB.88.094411}, is characterized by 
combination of sharp modes and a broader continuum arising from 
several multi-particle Majorana excitations. The lower boundaries  
of these excitations are defined by 
\begin{equation}\label{Eq:LowerBound}
\omega_{l}(q) \approx \sqrt{m^2_\text{thres} + v^2(q_x - q^\text{min}_x)}, 
\end{equation}
where $v = \frac{\pi J_\text{leg}}{2}$ is the spin velocity of the chain and  
$m_\text{thres}$ and ${\bf q}^\text{min}$ depend on the particles involved in the 
excitation. 
A summary of the relevant values can be found in Table I of 
Ref. \onlinecite{PhysRevB.88.094411}, which we have reproduced in 
Table II for convenience. From this table, one can see that a Majorana 
triplet (1T) excitation appears 
near ${\bf q}^\text{min} = (\pi/a,\pi/a)$ and $m_\text{thres} = m$, while 
the excitations near ${\bf q}^\text{min} = (\pi/a,0)$ correspond to a three-particle bound state 
consisting of two Majorana triplets and a Majorana singlet (2T + 1S) with a threshold 
set by $m_\text{thres} = 5m$.  
While the mapping to the Majorana fermion picture holds for $r \ll 1$, 
recent DMRG results for $S(q,\omega)$~\cite{PhysRevB.88.094411} have shown that this picture 
provides a qualitative description of the excitation spectrum for a wide range of $r  < 1$. 
These same calculations also showed that the spectral 
weight of the multi-particle continuum increases as $r \rightarrow 0$ until the  
entire excitation spectrum converges to the expected two-spinon continuum of the 
antiferromagnetic Heisenberg chain with lower and upper boundaries given by 
$\omega_s^{l}(q) = \frac{\pi}{2}J\left|\sin(qa)\right|$ and $\omega_s^{u}(q) = \pi
J\left|\sin(qa/2)\right|$, respectively.  

\begin{table}[h] 
	\begin{tabular}{p{2.0cm}p{2.0cm}p{2.0cm}l}
		\hline\hline
		Excitation & ${\bf q}^\mathrm{min}$ & $m_{thres}$   \\ \hline
		 1T       & $(\pi/a,\pi/a)$ & $1m$       \\ 
		 2T       & $(0,0)$         & $2m$     \\ 
		 3T       & $(\pi/a,\pi/a)$ & $3m$     \\ 
		 1T + 1S  & $(0,\pi/a)$     & $4m$       \\ 
		 2T + 1S  & $(\pi/a, 0)$    & $5m$   \\ 
		\hline\hline
	\end{tabular}\label{Tbl:Mthres}
	\caption{The momentum ${\bf q}^\mathrm{min}$ and $m_{thres}$ values that define the lower boundaries of 
        the single- and multiparticle excitations that occur in the Majorana fermion description of the  
        spin-$\frac{1}{2}$ ladders.  Reproduced from Ref. \onlinecite{PhysRevB.88.094411}.}            
\end{table}   

Understanding the behavior of a small number of holes doped into an antiferromagnetic background 
is one of the central problems in the quest to comprehend unconventional superconductivity.
In this context less is known about the excitations in doped spin-ladders as 
compared to the undoped case, where the former are usually
studied using numerical methods~\cite{PhysRevB.45.5744, PhysRevB.53.251, PhysRevB.94.155149, PhysRevLett.115.056401, PhysRevB.85.195103,PhysRevB.96.205120}. 
A single hole doped into a two-leg ladder introduces a spin $\frac{1}{2}$ and charge $+e$ to the system. 
DMRG results~\cite{PhysRevB.94.155149} indicate that in the strong-rung coupling limit, 
the doped hole behaves as a 
quasiparticle, where the spin and charge remain tightly bound within a typical distance of about one lattice constant. 
In the isotropic limit, the quasiparticle develops more internal structure with a length scale of $\sim 3a$. 
In the decoupled case ($J_\text{rung} = 0$), the doped hole 
fractionalizes completely into a spinon and holon~\cite{PhysRevLett.20.1445, KUMARNJP1}.

In the analysis below, we explore the RIXS spectra in both the NSC and SC channels 
and identify the  relevant elementary excitations in these spectra.

\begin{figure*}[!htbp]    
	\centering
	\begin{minipage}{0.49\linewidth}         
		\includegraphics[scale=0.95]{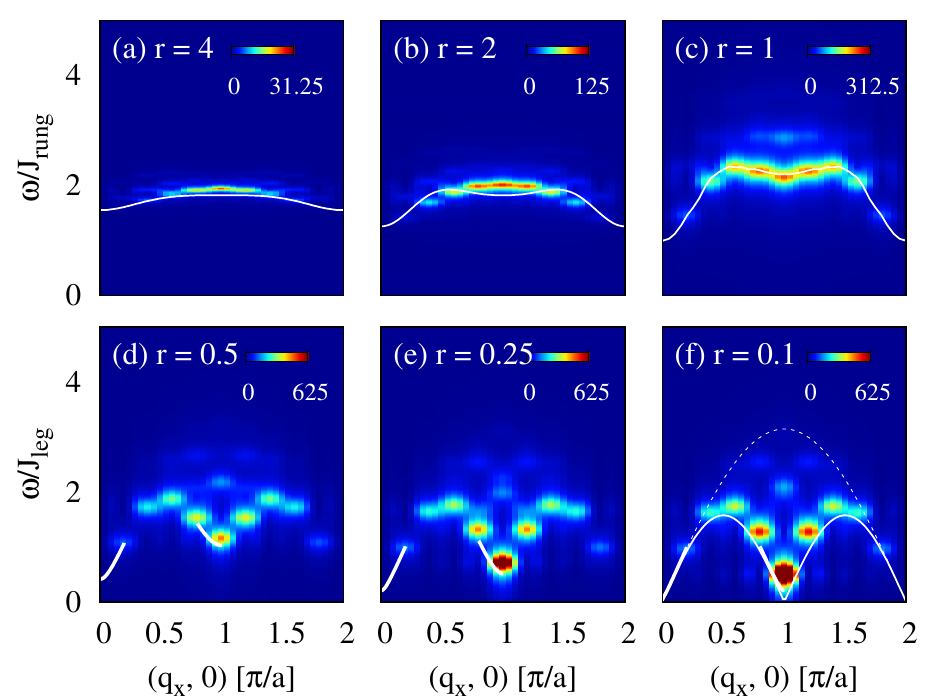}
	\end{minipage}
	\begin{minipage}{0.49\linewidth}
		\includegraphics[scale=0.95]{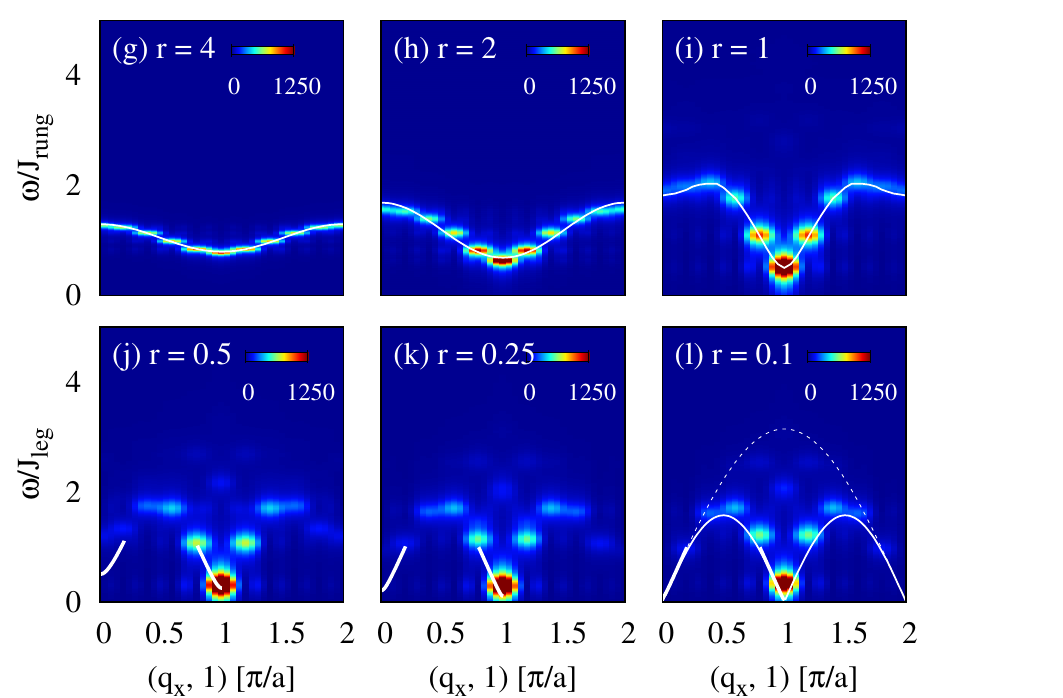}            
	\end{minipage}
	\caption{The RIXS spectra in the non-spin-conserving channel for a half-filled $t$-$J$ ladder, using ED and a 2$\times$10 cluster. 
		$I^{\Delta S=1}(q_x, 0, \omega)$ and $I^{\Delta S=1}(q_x, \pi/a, \omega)$ are shown in 
		panels (a)-(f) and (g)-(l), respectively. 
		Panels (a)-(c) and (g)-(i) have overlays of the dispersion relationships 
                for the bound \textit{triplet} two-triplon [Eq. (\ref{ttriplon})] and 
                the one-triplon excitations [Eq. (\ref{Eq:Triplon})], respectively, calculated using perturbation theory.  
                Panels (c)  and (i) have overlays (solid white) of the dispersion relationships for the same excitations 
                extracted from Ref. \onlinecite{doi:10.1142/S0217984905009237}, which were evaluated 
                using a continuous unitary transformation (CUT) method. 	
                The thin dotted and solid white lines in panels (f) and (l) plot the upper and lower 
                boundaries of the two spinon continuum expected for completely decoupled chains.  
                The thick white lines in panels (d)-(f) and (j)-(l) plot the lower boundaries of the multiparticle 
                continua near their respective minima. 
                Note that the $y$-axis of the top and botton rows are scaled with 
		respect to $J_\text{rung}$ and $J_\text{leg}$, respectively.}
	\label{fig:rixsundopedS1}
\end{figure*}

\subsection{Results for the non-spin-conserving channel\label{NSCchannel}}
We begin our study with the NSC or ``spin-flip" channel,  
which typically dominates the Cu L-edge RIXS spectra in cuprates~\cite{PhysRevX.6.021020, PhysRevLett.112.147401}.
The NSC channel produces local single spin-flips due to a large spin-orbit
coupling in the $2p$ core level, as shown in Fig~\ref{fig:schematicladder}(b). 
The elementary excitations generated in this scattering channel correspond to magnetic 
excitations with $\Delta S=1$ relative to the ground state.  
In terms of the spectra, the NSC channel is comparable
to the spin-flip channel of INS, and hence RIXS spectra at the Cu L-edge of
cuprates compare well with $S(\textbf{q}, \omega)$ 
(see Appendix~\ref{effectiveCorrelations})~\cite{Jia2014, PhysRevX.6.021020}.  

\subsubsection{Undoped $\boldmath{t}$-$\boldmath{J}$ ladders}	

The RIXS spectra in the NSC channel for undoped ladders are plotted in Fig.~\ref{fig:rixsundopedS1}. Panels (a)-(f) and (g)-(l) show results for 
momentum transfers ${\bf q} = (q_x, 0)$ and ${\bf q} = (q_x,\pi/a)$, respectively. Several excitations are identified.
	
In the limit of strong rung coupling, the spectra along the ${\bf q} = (q_x,0)$ direction 
[Figs.~\ref{fig:rixsundopedS1}(a) and \ref{fig:rixsundopedS1}(b)] exhibit a dispersive quasiparticle-like 
excitation. To determine its nature, we overlayed the dispersion $\omega^T_{2t}(q)$ given by Eq.~(\ref{ttriplon}). 
We find that the observed excitation closely follows the dispersion 
relationship for  $r  = 4$ but for $r = 2$ there are some
 deviations, most notably at the zone boundary. 
(The disagreement becomes even more apparent for $r = 1$, as discussed below.)  
The agreement between the
dispersion of the excitations and $\omega^T_{2t}(q)$, and the fact that we
are in the NSC channel, allows us to conclude that these excitations are the 
two-triplon bound state in the {\it triplet} channel. 
Similarily, the ${\bf q} = (q_x,\pi/a)$ excitation in the strong-rung coupling case  
[Figs.~\ref{fig:rixsundopedS1}(g) and \ref{fig:rixsundopedS1}(h)] corresponds to a single 
triplon excitation. To confirm this, we overlayed the dispersion $\omega_t(q)$ 
given by Eq. (\ref{Eq:Triplon}), showing it captures well the observed excitations for $r\geq 2$.    
	
As discussed in the previous section, in the weak-rung coupling limit we expect the legs of the ladders to 
behave as weakly coupled antiferromagnetic chains. Indeed, along both the
$(q_x, 0)$ [Figs.~\ref{fig:rixsundopedS1}(d)-(f)] and
$(q_x,\pi/a)$ [Figs.~\ref{fig:rixsundopedS1}(j)-(l)]
directions, the spectra can be described using the picture of confined spinons with a 
continuum of excitations appearing above a sharper dispersing mode. 
The lower boundaries of the continua near their respective minima are overlaid as thick white 
lines. 
According to Table II, we assign the excitations near ${\bf q}^\text{min} = (0,0)$ to 2T 
excitations with $m_\text{thres} = 2m$, while the excitations near ${\bf q}^\text{min} = (\pi/a,0)$ 
correspond the 2T + 1S excitations with  $m_\text{thres} = 5m$.    
Similarily, the excitations near ${\bf q}^\text{min} = (\pi/a,\pi/a)$ are 1T 
excitations with $m_\text{thres} = m$ and the excitations near ${\bf q}^\text{min} = (0,\pi/a)$ 
correspond to 1T + 1S excitations with $m_\text{thres} \approx 4m$. 
	
The \textit{isotropic} coupling case behaves qualitatively like the strong-rung coupling cases, 
but the calculated spectra deviate significantly from the dispersion predicted by perturbation theory [Eq. (\ref{ttriplon})]. 
Nevertheless, we are still able to assign the intense dispersing features to the $S = 1$ two-triplon bound state and 
the one-triplon excitations, as in the strong-rung coupling limit.  
In Fig.~\ref{fig:rixsundopedS1}(c) and Fig.~\ref{fig:rixsundopedS1}(i) 
we have overlayed the dispersions for the bound {\sl triplet} two-triplon and one-triplon  
excitations, respectively, this time extracted from Fig. 4 of Ref.~\onlinecite{doi:10.1142/S0217984905009237}. 
In this case, the dispersions were computed using a continuous unitary transformation (CUT) method, 
and agree well with our evaluated spectra. In Fig.~\ref{fig:rixsundopedS1}(c) we also observe additional 
spectral weight at higher energies, which corresponds to unbound excitations inside the two-triplon continuum. 
Our results in this regime should be of considerable interest for future RIXS experiments 
on cuprates spin-$\frac{1}{2}$ ladder materials, as most of the estimated values of 
the $J_\text{rung}/J_\text{leg}$ ratios fall in this intermediate category.

\begin{figure*}[ht]    
	\centering
	\begin{minipage}{0.49\linewidth}         
		\includegraphics[scale=0.95]{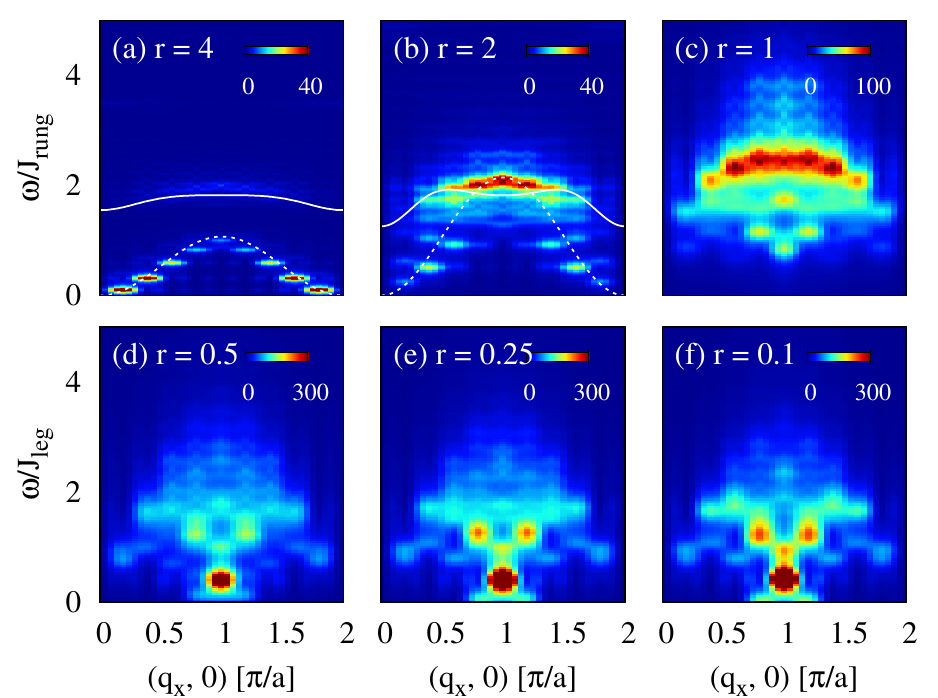}
	\end{minipage}
	\begin{minipage}{0.49\linewidth}
		\includegraphics[scale=0.95]{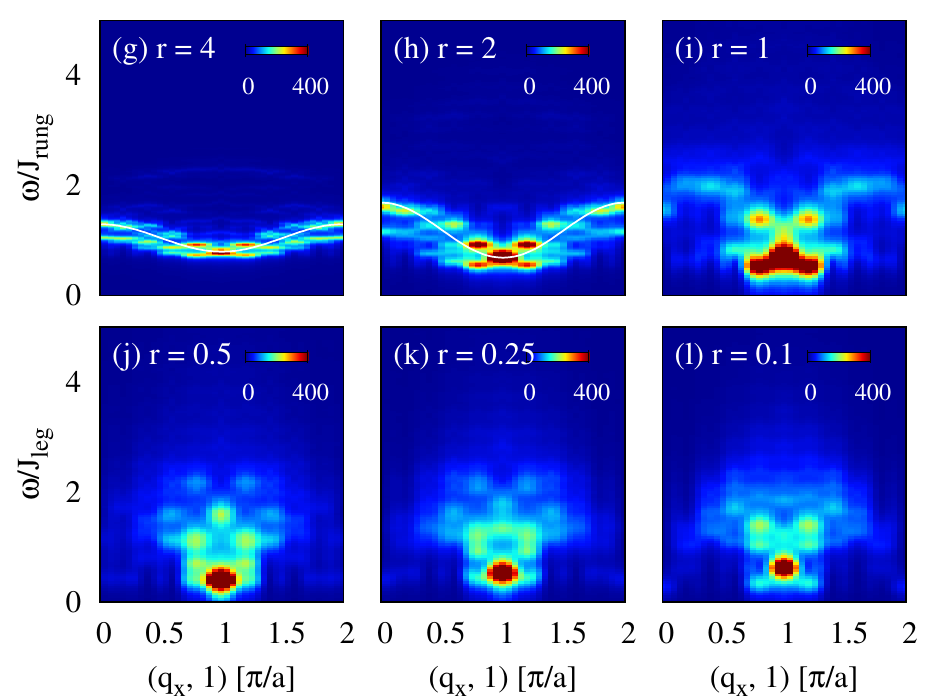}
	\end{minipage}    
	\caption{RIXS spectra in the non-spin-conserving channel for a doped $t$-$J$ ladder, using ED on a 2$\times$10 cluster and a filling of $\langle n\rangle=0.9$. 
		$I^{\Delta S=1}(q_x, 0, \omega)$ and $I^{\Delta S=1}(q_x, \pi/a, \omega)$ are in panels (a)-(f) and (g)-(l), respectively. 
		Panels (a), (b) and (g), (h) have overlays (solid white) 
                of the dispersion relationships for the bound \textit{triplet} two-triplon [Eq. (\ref{ttriplon})] and 
                one-triplon excitations [Eq. (\ref{Eq:Triplon})], respectively, derived using perturbation theory. 
                Panels (a) and (b)
		have an additional overly (dashed white) of the dispersion for a quasiparticle 
               $\omega(k) = 2\tilde{t}[1-\cos(ka)]$. 
		The $y$-axis of the top and botton rows are scaled with respect to $J_\text{rung}$ and $J_\text{leg}$, respectively.}
	\label{fig:rixsdopedS1}    
\end{figure*}

\subsubsection{Doped $t$-$J$ ladders}

The excitations of doped ladder compounds are 
relevant to explain pressure-induced superconductivity. 
Moreover, while low-energy spin-fluctuations are widely considered pivotal for superconductivity, 
the relationship between the doping evolution of charge and high-energy spin excitations and the 
superconducting mechanism has recently become the subject of considerable debate, especially in 
2D cuprates. Our results for the RIXS spectra of the doped spin-ladder in the NSC-channel are in
Fig.~\ref{fig:rixsdopedS1}. As in the undoped case, panels (a)-(f) and (g)-(l) show spectra along the
($q_x, 0$) and ($q_x,\pi/a$) directions, respectively. 
	
For strong-rung couplings [Figs.~\ref{fig:rixsdopedS1}(a) and \ref{fig:rixsdopedS1}(b)], 
the spectra along the ${\bf q} = (q_x, 0)$ directions have two distinct sets of 
excitations. The first corresponds to the same {\it triplet} two-triplon excitations identified 
in the undoped case, as confirmed by overlaying the dispersion given by 
Eq.~(\ref{ttriplon}) as solid white lines.  
The second is the Bloch quasiparticle excitation formed from the tightly bound 
spin and charge of the doped hole~\cite{PhysRevB.94.155149}.   
Its dispersion is well described by $\omega(k) = 2\tilde{t}[1-\cos(ka)]$ (the dashed line overlay), where 
$\tilde{t}=t_\text{rung}/2$ is the effective hopping of a quasiparticle 
in the bonding band~\cite{PhysRevB.53.251}. 
The fact that the spectra exhibits gapless charge and gapped spin (C1S0) excitations is consistent 
with the system's classification as a Luther-Emery liquid~\cite{PhysRevB.58.3425, PhysRevB.53.251}.
The spectra along the ${\bf q} = (q_x, \pi/a)$ direction, shown in
Figs.~\ref{fig:rixsdopedS1}(g) and ~\ref{fig:rixsdopedS1}(h), have only a single
set of excitations, whose dispersions agree well with the
one-triplon excitation Eq.~(\ref{Eq:Triplon}), which is again
overlaided as a solid white line. 

Results for the weak-rung coupling regime along the ${\bf q} = (q_x, 0)$ and $(q_x, \pi/a)$ 
directions are shown in Figs.~\ref{fig:rixsdopedS1}(d)-(f) 
and Figs.~\ref{fig:rixsdopedS1}(j)-(l), respectively.  
We find that the spectra soften as compared to undoped spin ladders. 
Moreover, the spin gap no longer appears to scale with $J_\text{rung}$ but instead 
appears to vanish at $(\pi/a, 0)$ for all $r < 1$ 
while persisting at ${\bf q} = (\pi/a, \pi/a)$.

\begin{figure*}[!ht]    
	\centering
	\begin{minipage}{0.49\linewidth}         
		\includegraphics[scale=0.95]{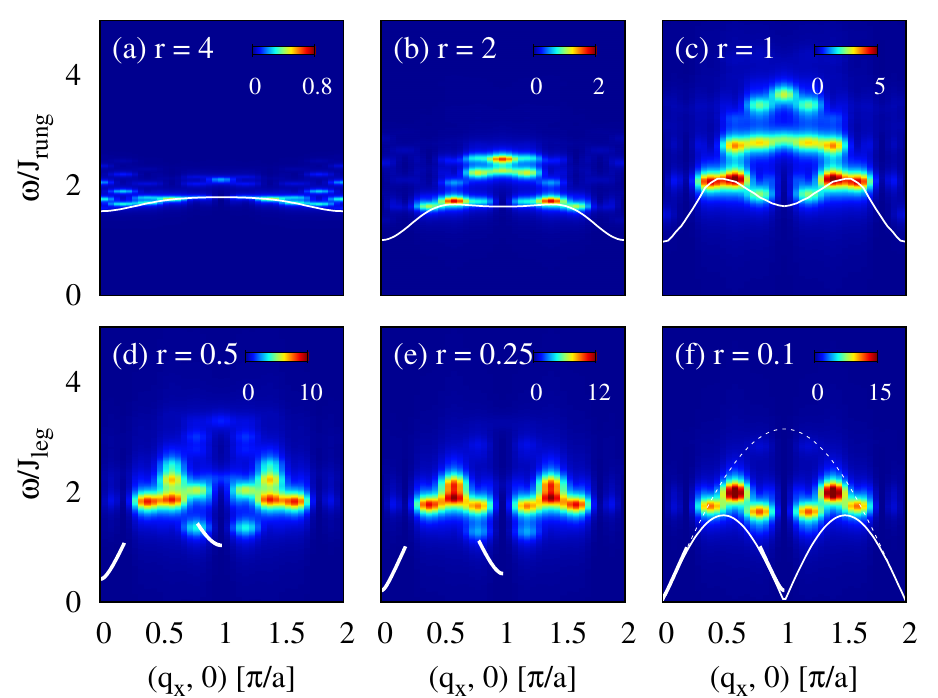}
	\end{minipage}
	\begin{minipage}{0.49\linewidth}
		\includegraphics[scale=0.95]{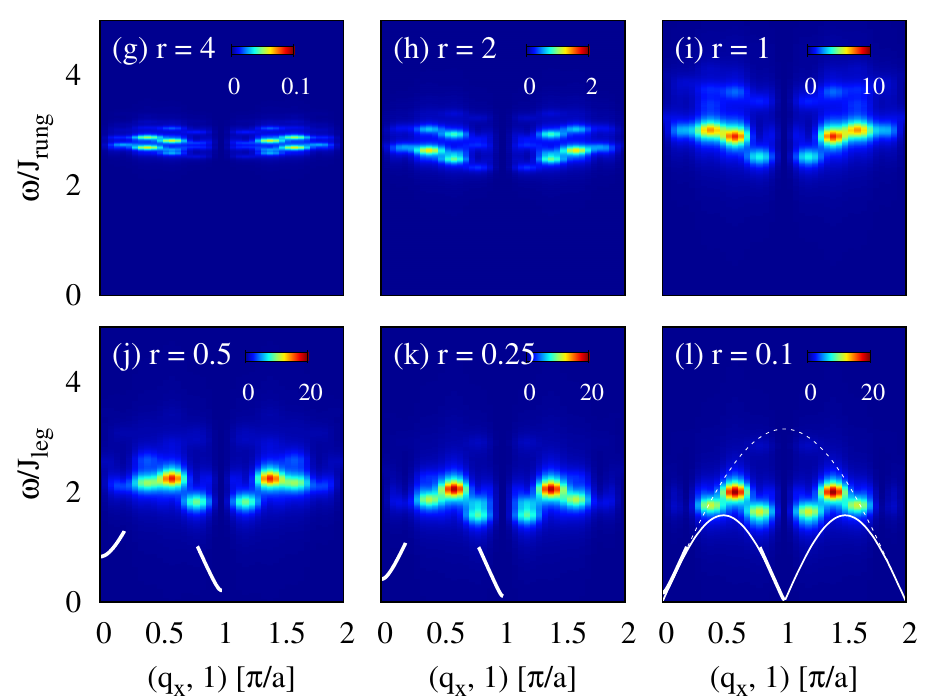}
	\end{minipage}    
	\caption{RIXS spectra in the spin-conserving channel for the half-filled $t$-$J$ ladder, using ED and a 2$\times$10 cluster. 
		$I^{\Delta S=0}(q_x, 0, \omega)$ and $I^{\Delta S=0}(q_x, \pi/a, \omega)$ are in panels 
		(a)-(f) and (g)-(l), respectively. Panels (a) and (b) have overlays (solid white) of the dispersion relations
                of the \textit{singlet} bound two-triplon excitations [Eq. (\ref{striplon})] derived using perturbation theory, 
                while panel (c) has an overlay of the dispersion relation for the same excitation 
                from Ref. \onlinecite{doi:10.1142/S0217984905009237}, using the CUT method. 
                The thin dotted and solid white lines in panels (f) and (l) plot the upper and lower 
                boundaries of the two spinon continuum expected for completely decoupled chains. 
                The thick white lines in panels (d)-(f) and (j)-(l) plot the lower boundaries of the multiparticle 
                continua near their respective minima.  
                The $y$-axis of the top and botton rows are plotted in units of $J_\text{rung}$ 
		and $J_\text{leg}$, respectively.}
	\label{fig:rixsundopedS0}
\end{figure*}

In the \textit{isotropic} case at ${\bf q} = (0, \pi/a)$, shown in Fig.~\ref{fig:rixsdopedS1}(c), 
the brightest dispersing peak does not have the same downturn in the two-triplon dispersion 
that was observed in the undoped case. Instead, there is an increased weight appearing in at higher 
energy losses, corresponding to the two-triplon continuum. 
In contrast, the excitations at ${\bf q} = (\pi/a, \pi/a)$, 
shown in Fig.~\ref{fig:rixsdopedS1}(i), have the ubiquitous incommensurate peaks that 
are also commonly observed in doped ladders and 2D
cuprates~\cite{PhysRevB.96.205120, PhysRevB.97.235137, Tranquada2004, Xu2009}.
It is interesting to contrast the results for the doped two-leg spin ladder found above with 
available results in the doped 2D cuprates at the Cu L-edge. In 2D cuprates, a weakly dispersive 
high-energy paramagnon band along the ${\bf q} = (q_x, 0)$ line was reported to be persistent 
upon hole doping~\cite{LeTacon2011,Jia2014,PhysRevB.88.020501,PhysRevLett.114.217003,Huang2016}. 
This type of excitation compares relatively well with our results in the two-leg ladder case 
in Fig.~\ref{fig:rixsdopedS1}(c) for the isotropic case.

\subsection{Results for the spin-conserving channel\label{SCchannel}}
We now analyze the RIXS spectra in the SC channel, both for the undoped and doped cases.  As shown pictorially in 
Figs.~\ref{fig:schematicladder}(c) 
and ~\ref{fig:schematicladder}(d), the magnetic excitations that are accessible in this channel 
occur via double spin-flip processes, which correspond to $\Delta S = 0$ excitations in the antiferromagnetic ladders. 
For the undoped cuprates measured at the 
Cu $L$-edge, the SC channel probes excitations encoded in the dynamical exchange structure factor 
$S^\text{exch}(\textbf{q}, \omega)$ (see Appendix~\ref{effectiveCorrelations}), which is a 
second order term in the ultrashort core-hole lifetime 
(UCL) expansion~\cite{PhysRevB.83.245133, PhysRevX.6.021020}.
Because these are higher order processes, this channel is expected to be weaker as 
compared to the NSC channel,~\cite{PhysRevLett.112.147401, PhysRevX.6.021020} and our results are consistent with this expectation. 
In the doped case, magnetic and charge excitations coexist in the RIXS 
spectra and the SC channel also has a significant contribution at second order given 
by a modified charge structure factor  $\tilde{N}(\textbf{q},\omega)$ 
[see Eq. (9) of Ref.~\onlinecite{PhysRevX.6.021020} and Appendix~\ref{effectiveCorrelations}].	
The SC channel is also particularly relevant at the O and Cu K-edges, where direct 
spin-flip excitations are often forbidden~\cite{PhysRevB.85.214527, KUMARNJP1, FSDIR}.
Our numerical study motivates RIXS experiments that could be able to disentangle SC and NSC components 
of the spectra by the use of photon polarization, which has been successfully demonstrated 
in Ref.~\onlinecite{PhysRevLett.112.147401} for the weakly coupled ladder cuprate 
CaCu\textsubscript{2}O\textsubscript{3}.
	
\subsubsection{Undoped $t$-$J$ ladders}

The RIXS spectra in the SC channel for the undoped ladders are shown in Fig.~\ref{fig:rixsundopedS0}. 
Panels (a)-(f) and (g)-(l) show the RIXS spectra for momentum transfers 
along the ${\bf q} = (q_x, 0)$ and $(q_x,\pi/a)$ directions, respectively. 
As expected, the intensity of the excitations in this channel is weaker as compared to  the
NSC channel by approximately one order of magnitude.

As with the previous sections, we first consider the strong-rung coupling limit.  	
Along the $(q_x, 0)$ direction [Figs.~\ref{fig:rixsundopedS0}(a) and \ref{fig:rixsundopedS0}(b)], 
we observe a weakly dispersing feature that agrees well with the two-triplon bound state in the 
{\it singlet} channel given by Eq. (\ref{striplon}). This is one of the
important results of our current investigation: because the SC channel
probes $\Delta S=0$ excitations, we are able to clearly identify and distinguish the
two-triplon bound states in both the singlet and triplet channels. We also see additional spectral
weight at higher binding energies near $(\pi/a,0)$, which contrasts qualitatively with the results in the
weak-rung coupling limit. This weight falls within the two-triplon continuum
and likely corresponds to unbound two-triplon excitations in the singlet
channel. 

Along the ${\bf q} = (q_x, \pi/a)$ direction in the strong-rung coupling regime
shown in panels (g) and (h), the spectra can again be understood in 
terms of {\sl singlet} two-triplon excitations, which we observe 
at energy losses around 3$J_\text{rung}$. We observe zero spectral weight at
$(\pi/a, \pi/a)$ for all the rung couplings 
we investigated, in contrast to the weak spectral weight observed at $(\pi/a, 0)$. 
The nature of these excitations can be qualitatively captured using the
dynamical exchange structure factor $S^\text{exch}(\textbf{q},\omega)$, which
has been computed and shown in Fig.~\ref{fig:S_exchS0} of 
Appendix~\ref{effectiveCorrelations}. Indeed, 
$S^\text{exch}(\textbf{q}, \omega)$ shows the spectral weight
cancellation  at $(\pi/a, \pi/a)$ for all the rung couplings investigated.
\begin{figure*}[!ht]    
	\centering
	\begin{minipage}{0.49\linewidth}         
		\includegraphics[scale=0.95]{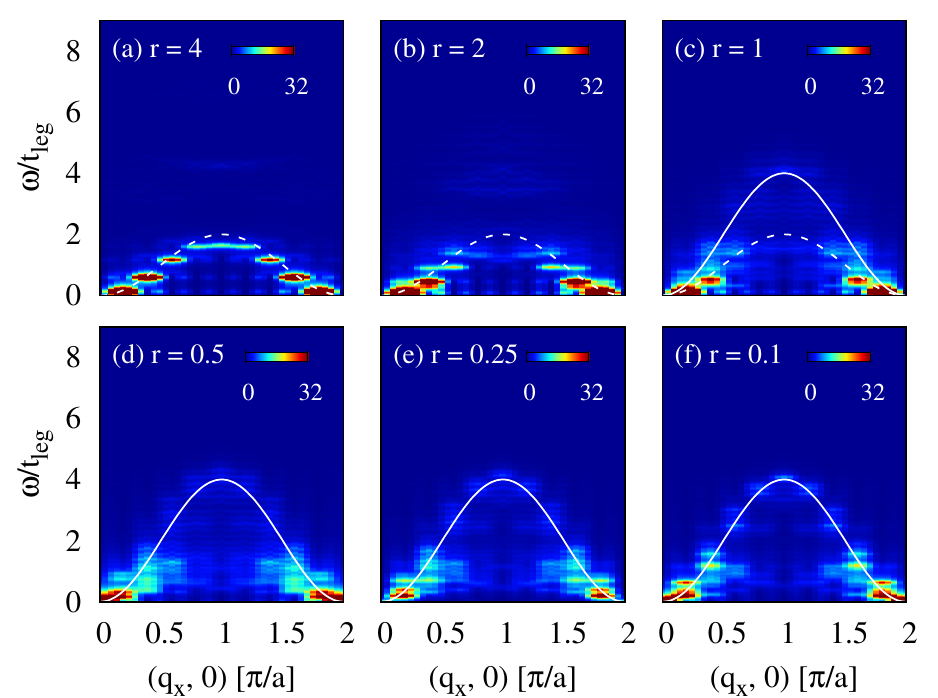}
	\end{minipage}
	\begin{minipage}{0.49\linewidth}
		\includegraphics[scale=0.95]{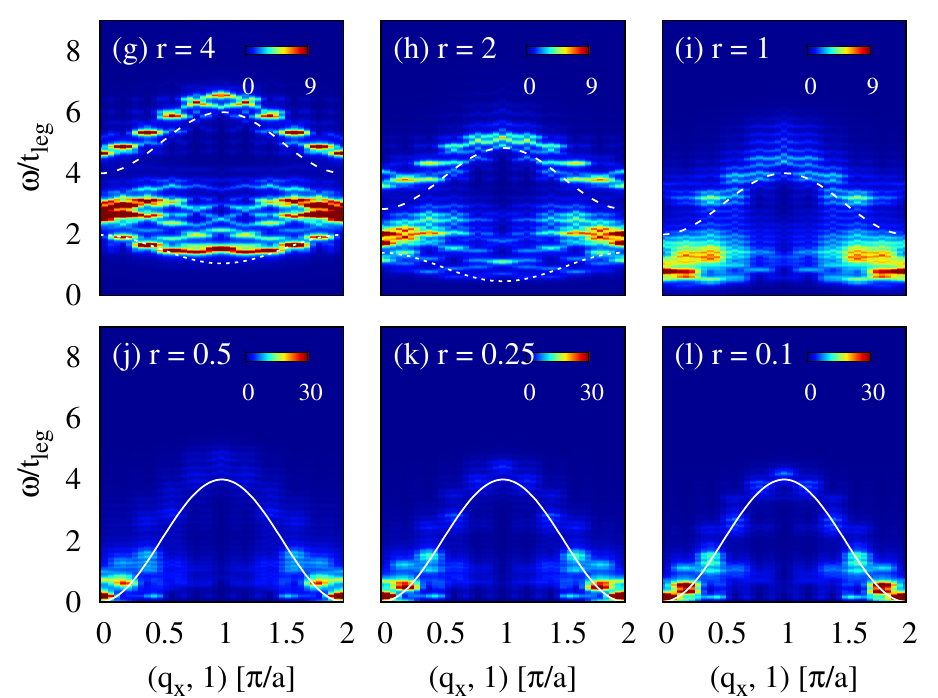}
	\end{minipage}    
	\caption{RIXS spectra in the spin-conserving channel for the doped $t$-$J$ ladder, using ED, a 2$\times$10 cluster, and $n=0.9$.
		$I^{\Delta S=0}(q_x, 0, \omega)$ and $I^{\Delta S=0}(q_x, \pi/a,\omega)$ are in panels (a)-(f) and (g)-(l), 
		respectively. In panels (a)-(c) the dispersions of the quasiparticle 
		state with a bandwidth $W = 2t_\text{rung}$ are shown  (white dashed). In panels (d)-(f) and (j)-(l) 
		the holon dispersion with a bandwidth $W = 4t_\text{leg}$ is also shown (solid white).  The white dashed 
                overlay in panels (g)-(i) 
		plot the boundary of the spinon-holon continuum gapped by $2t_\text{rung}$, while the 
		dotted overlay in panels (g) and (h) corresponds to the dispersion relation of the one-triplon excitations. 
		The $y$-axis of the top row and botton rows 
		are plotted in units of $t_\text{leg}$.}
	\label{fig:rixsdopedS0}
\end{figure*}

In the weak rung regime ($r < 1$), the spectra along ${\bf q} = (q_x, 0)$
and $(q_x, \pi/a)$, shown in
Figs.~\ref{fig:rixsundopedS0}(d)-(f)
and~\ref{fig:rixsundopedS0}(j)-(l), respectively,
resemble the continuum expected for confined spinons also observed in the NSC
channel. To highlight this, we overlaid the boundaries of the 
two-spinon continuum as well as the lower boundaries of the 
multi-particle continua that were introduced when 
describing the NSC channel. 
In this case, all of the excitations appear above the lower boundary lines, indicating that these excitations are multiparticle in nature. 
We also note
that spectra along both momentum directions have a suppressed intensity at $q_x
=\pi/a$, which is similar to what occurs in one-dimensional antiferromagnetic
chains when probed in  the SC channel~\cite{PhysRevB.85.064423,
PhysRevB.77.134428, PhysRevLett.106.157205}.

The spectra along both momentum directions for the isotropic case ($r = 1$) 
behaves qualitatively similar to the strong-rung coupling case, 
where we observe a continuum of excitations. In
Fig.~\ref{fig:rixsundopedS0}c, we plot an overlay 
extracted from Fig. 4(b) of
Ref.~\onlinecite{doi:10.1142/S0217984905009237} for bound
two-triplon excitation in the {\it singlet} channel, again evaluated using the CUT method. 
This dispersion agrees well with the lower boundary
of the evaluated spectra suggesting that the continuum of excitations is related to {\it singlet} two-triplon excitations 
and that the bound singlet state is not far removed from the continuum. 	
	
\subsubsection{Doped $t$-$J$ ladders}
Finally, we examine the RIXS spectra of the doped ladders in the SC channel.  
As before, Figs.~\ref{fig:rixsdopedS0}(a)-(f) and \ref{fig:rixsdopedS0}(g)-(l) 
show results for momentum transfers  ${\bf q} = (q_x, 0)$ and $(q_x,\pi/a)$, respectively. 
The spectra are quite rich and we observe several new excitations that were not
present in the undoped ladders. Indeed, we expect that
magnetic and charge excitations coexist in the SC channel response 
and that most of the spectral features observed in the full RIXS response can be described 
using the modified dynamical charge correlation function 
$\tilde{N}(\textbf{q},\omega)$ (see Fig.~\ref{fig:nqwdoped} in Appendix~\ref{effectiveCorrelations} 
and Ref. \onlinecite{PhysRevX.6.021020}). 
	
\begin{figure}[h]
	\centering
	\includegraphics[width=\linewidth]{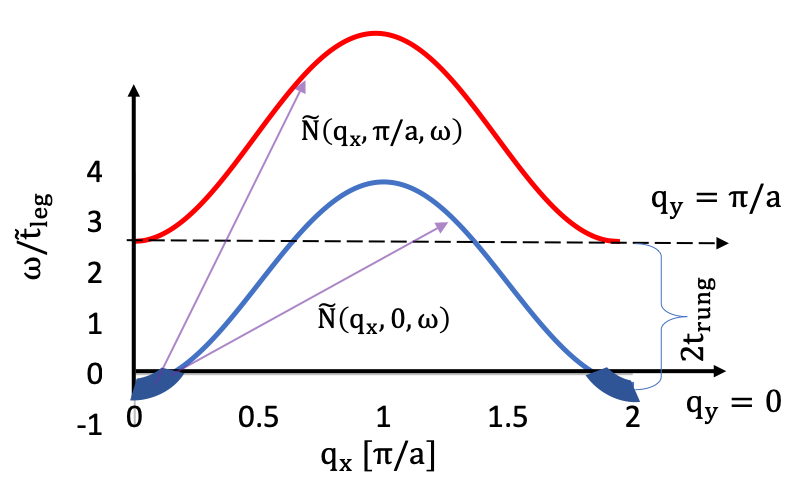}
	\caption{A schematic diagram of the particle-hole excitations possible in the doped $t$-$J$ 
		ladder in the weak-rung coupling regime. In this limit, the orbitals 
                along the legs form  
		bonding ~$\tfrac{1}{\sqrt{2}}(c_{i,\sigma, 1}^\dagger-c_{i,\sigma,2}^\dagger)|0\rangle$ 
		and anti-bonding  
		~$\tfrac{1}{\sqrt{2}}(c_{i,\sigma,1}^\dagger+c_{i,\sigma,2}^\dagger)|0\rangle$ states, 
                which form the basis for the Bloch states propagating along the leg direction. 
                In this case, the dispersions of the two Bloch states are split by an amount proportional to the rung hopping $\pm t_\text{rung}$. The doped holes 
                form quasiparticles carrying both spin-$\frac{1}{2}$ and charge $e$, and the broad blue colour highlights the 
		filled states in the ground state. RIXS probes scattering within the bonding and to
		the antibonding band, as shown by the arrows.}
	\label{fig:dopedexcitations}
\end{figure}

As before, we begin our discussion from the strong-rung coupling limit. 
In this case, the ladder can be considered as composed of weakly decoupled dimers where the orbitals on each leg form bonding and antibonding states. 
If $t_\text{leg}$ is finite, these bonding (-) and antibonding (+) states form the basis 
for Bloch states with dispersion relations given 
by $\omega(q)=\mp t_\text{rung}+2\tilde{t}[1-\cos(qa)] $~\cite{PhysRevB.53.251}, 
where $\tilde{t}$ is the effective hopping parameter obtained 
from the change of basis to the bonding and antibonding states. 
When a small number of holes are doped into the system, they first occupy the bonding band as quasiparticles, as shown in Fig.~\ref{fig:dopedexcitations}. 
The charge excitations observed in this channel can then be understood
by invoking quasiparticle scattering within and between the bonding and antibonding bands, respectively.

Along the $(q_x, 0)$ direction in the strong-rung coupling regime, Figs.~\ref{fig:rixsdopedS0}(a) and 
\ref{fig:rixsdopedS0}(b), we observe dispersive charge excitations 
consistent with particle-hole scattering within the bonding band, as
shown in Fig.~\ref{fig:dopedexcitations}. To confirm this, we overlaid  
the dispersion $\omega(q) =2\tilde{t}[1-\cos(qa)]$, where
$\tilde{t} \approx t_\text{leg}/2$, which agrees with the numerical data~\cite{PhysRevB.53.251}.      	
Along the $(q_x, \pi/a)$ direction [Figs.~\ref{fig:rixsdopedS0}(g) and \ref{fig:rixsdopedS0}(h)]  
we find the corresponding particle-hole excitation where scattering occurs into the
antibonding band. In this case, we overlaid the dispersion 
$\omega(q)=2t_\text{rung}+2\tilde{t}[1-\cos(qa)]$ (white dashed line). 
We also notice that the bonding and anti-bonding bands are separated by $2t_\text{rung}$, 
which accounts for the shift in cosine-like dispersion observed when $q_y = \pi/a$. 

It is important to note that, even in this case, these charge excitations are weaker in intensity when 
compared to the magnetic excitations in NSC channel by approximately one order of magnitude, 
but stronger than the SC channel of the undoped case. 
Our results show that  RIXS can
explicitly probe charge excitations at low energies as suggested in
the literature for the Cu $L_3$-edge~\cite{PhysRevLett.112.247002,PhysRevB.94.165127}.

In addition to the charge excitations, we also observe a continuum of magnetic excitations for momentum transfers 
along $(q_x,\pi/a)$. The lower boundary of this continuum is defined by the one triplon dispersion given by 
Eq.~(\ref{Eq:Triplon}), which has been overlaid as a dotted white line. 

In the weak-rung coupling regime, shown in 
Figs.~\ref{fig:dopedexcitations}(d)-(f) and 
\ref{fig:dopedexcitations}(j)-(l), the quasiparticle excitations 
display a bandwidth $4t_\text{leg}$ with dispersion 
$\omega(q) = 2t_\text{leg}[1-\cos(qa)]$ along both the $(q_x,0)$ and $(q_x,\pi/a)$ directions. 
This is similar to the results for the 1D AFM chain reported in Ref.~\onlinecite{KUMARNJP1}  where 
the excitations are holons, and consistent 
with the notion that the individual legs of the ladder are weakly coupled.  The fact that these observed excitations 
are completely governed by $t_\text{leg}$ indicates that the holes occupy the chains of the ladder rather than the 
bonding and antibonding orbitals on each rung. 	

Finally, we consider the isotropic rung-coupling limit, which is 
of much interest for future RIXS experiments and future theoretical investigations.
In this case, our spectra show a gapless and a gapped continuum along the $(q_x,0)$ 
and $(q_x,\pi/a)$ directions, respectively. These results hence resemble qualitatively the spectral features 
also observed in the strong-rung coupling limit. It is interesting to 
compare our full RIXS spectra with the dynamical charge structure factor results reported in 
Fig. 7 of Ref.~\onlinecite{PhysRevB.97.195156}: our results compare well with the lower 
Hubbard band excitation observed in which $N(q_x, \pi/a, \omega)$ is
gapped, in contrast to the gapless $N(q_x, 0, \omega)$.

\subsection{Revisiting \SCO~RIXS data}\label{Schlappadata}
\begin{figure}[h]
\centering
\includegraphics[width=0.8\linewidth]{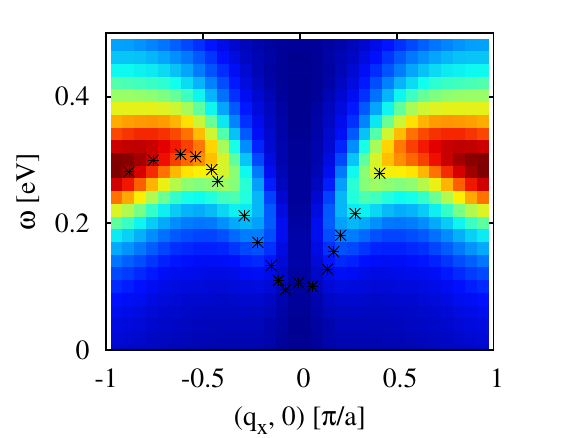}
\caption{
Calculated RIXS spectra of an undoped $t$-$J$ ladder in the non-spin-conserving 
channel $I^{\Delta S=1}(q_x, 0, \omega)$ 
evaluated using DMRG on an $N=16\times 2$ lattice. The black squares overlay the peak positions 
of the spectra extracted from the experimental Cu L$_3$-edge data for \SCO, reproduced 
from Ref.~\onlinecite{PhysRevLett.103.047401}.}
\label{fig:srcuoladder}
\end{figure} 

Cu $L_3$-edge RIXS data has been reported~\cite{PhysRevLett.103.047401} for the prototypical spin-ladder \SCO.     
At that time, the observed spectra were 
interpreted in terms of two-triplon $\Delta S = 0$ excitations in the strong rung coupling regime ($r \approx 1.37$) as it 
was believed that $\Delta S=1$ excitations were forbidden at the Cu $L_3$-edge. 
It was later shown that not only is this channel allowed but that it dominates the magnetic RIXS
response in the undoped cuprates~\cite{PhysRevLett.103.117003}. 
The RIXS spectra were later theoretically evaluated~\cite{PhysRevB.85.224436} employing a projection method 
for the two-leg ladder using the parameter set derived from La$_4$Sr$_{10}$Cu$_{24}$O$_{41}$ in Ref.~\onlinecite{PhysRevLett.98.027403}
(the model involved additional ring spin-exchange term as compared to our model).
These calculations showed that the RIXS spectra was associated with the two-triplon excitations 
with $\Delta S=1$ when momentum transfers of $q_y = 0$ were used, as shown in Fig. 6 of
Ref.~\onlinecite{PhysRevB.85.224436}. But there was also a significant difference in the dispersion of the experimental and theoretical data.  
Armed now with these theoretical insights and our new model calculations, we revisit the existing  \SCO~data.  
	
To access large system sizes with improved momentum resolution, 
we computed the RIXS spectra in the $\Delta S = 1$ channel for an undoped spin-ladder 
using our recently formulated DMRG approach~\cite{NoceraSR1}. 
To obtain a unified description of RIXS and INS experiments, we first  
adopted a model given by Eq.~(\ref{tJmodel}) without ring exchange terms and parameters from Ref.~\cite{PhysRevLett.81.1702}, 
however, we found  that this model gave poor agreement with the experimentally observed RIXS spectra. 
Instead, we are able to find good agreement when we set $J_\text{leg} = 145$ meV 
and $J_\text{rung} = 0.85J_\text{leg}$. This result places \SCO~in the weak rung-coupling  
regime but close to the isotropic limit. 
The resulting RIXS spectra shown in Fig.~\ref{fig:srcuoladder} agree well with
the experimental data, but the theoretical model predicts a vanishing spectral weight at $q_x=0$. 
This observation is consistent with the model of Ref.
\onlinecite{PhysRevB.85.224436} but inconsistent with the finite intensity
observed in the experiment.  At this time, the source of this discrepancy is
unclear. 

Because $r$ is close to the isotropic case, the data can be
qualitatively understood using either the dimer excitation picture or the
confined spinon picture. In the former case, the excitations are understood as
bound $S = 1$ two-triplet excitations. In the latter case, they are viewed as a
three particle bound state composed of Majorana fermions near the zone boundary. However,  a
quantitative description of the data can only be achieved with nonperturbative
numerical methods. In this sense, our results place \SCO~in a regime similar to
the organometallic compound
(C$_7$H$_{10}$N)$_2$CuBr$_4$~\cite{PhysRevB.88.094411}, but with larger
exchange couplings.
	
\section{Conclusions \label{conclusion}}

We have systematically studied the RIXS spectra of both undoped and doped 
spin-$\frac{1}{2}$ ladders, covering the weak- to strong-rung coupling regimes. Our study shows that  
RIXS experiments performed on these compounds can access a wealth of magnetic and charge excitations.  
This study was motivated by RIXS experiments at the Cu L-edge in low-dimensional cuprates, where the RIXS data
can be decomposed into the non-spin-conserving (NSC) and spin conserving 
(SC) channels~\cite{PhysRevB.85.064423, PhysRevLett.112.147401, PhysRevX.6.021020, PhysRevB.85.214527, PhysRevLett.112.147401}. Therefore, we evaluated the RIXS spectra in both of 
these channels and provided an energy-momentum resolved roadmap that can guide future RIXS experiments on spin-ladder compounds. 

In the first part of our effort, we reported the RIXS excitations in the NSC or ``spin-flip" channel, 
which typically dominates the Cu L-edge RIXS spectra in the cuprates~\cite{PhysRevX.6.021020, PhysRevLett.112.147401}. 
In the undoped two-leg ladder, we have shown that RIXS can access dispersive one triplon excitations and a 
two-triplon bound state in the {\sl triplet} ($S = 1$) channel in the intermediate to strong rung coupling regime. 
In the weak-rung coupling regime, the NSC channel probes single- and multiparticle excitations 
consistent with the Majorana fermion description of confined spinons. 
  
The study of the RIXS spectra for doped spin-$\frac{1}{2}$ ladder compounds is of much
importance in the context of pressure-induced superconductivity in
low-dimensional high-T\textsubscript{c} cuprates.  In the doped ladder, we
accessed  one- and (triplet) two-triplon excitations in the strong-rung coupling limit and
softened confined spinons in the weak-rung coupling limit. We also 
identified signatures of a bound spin-charge quasiparticle excitation in the
strong-rung coupling limit. 

In the second part of our work, we studied the RIXS spectra of the spin-ladder
in the SC channel, which probes $\Delta S=0$ excitations
of the system~\cite{PhysRevLett.112.147401, PhysRevX.6.021020}. This component
of the RIXS spectra has received less attention in the literature, 
and our work provides a starting point for future theoretical and
experimental investigations of this channel on spin-ladders. In the undoped 
ladders, magnetic excitations are created in this channel via double spin-flip
processes. Because these are higher order processes, their 
contribution to the RIXS spectra is expected to be weaker compared to the NSC
channel~\cite{PhysRevLett.112.147401, PhysRevX.6.021020}. Our results are
consistent with this expectation, and we found that the spectral intensity is
at least one order of magnitude smaller than the corresponding spectra in the
SC channel. Nevertheless, in the intermediate to strong rung coupling
regime, we are able to identify bound two-triplon excitations in the
singlet ($S=0$) channel.  

In the SC RIXS channel for doped spin-ladders, we identified a
set of dispersive low-energy charge excitations that 
were interpreted by invoking quasiparticle scattering within and between the
bonding and antibonding bands, respectively, in the strong rung coupling case. 
Conversely, the spectra are dominated by holon excitations in the
weak-rung coupling limit. The direct access to charge excitations offered 
by this channel provides a new opportunity to study superconductivity in
cuprate ladders, where the role of spin and charge excitations is still debated. 
We believe our numerical study   
motivates new RIXS experiments on spin-ladders such as \SCO~allowing
disentanglement of data into the NSC and SC channels, due to the richness predicted in
the RIXS spectra.

Finally, we revisited the available RIXS data for \SCO 
and found that it was best described using a model in 
the weak rung coupling regime with $r = 0.85$. This result is in contrast to the 
previous analysis~\cite{PhysRevLett.103.047401} that placed it in the strong rung
coupling regime but in qualitative agreement with the INS data~\cite{PhysRevLett.81.1702}. .  

\begin{acknowledgements}
We thank T. Schmitt and J. Schlappa for useful discussions.  A.~N. and E.~D.
are supported by the U.S. Department of Energy, Office of Science,  Basic
Energy Sciences, Materials Sciences and Engineering Division. 
S.~J. is supported by the National Science Foundation under Grant No.
DMR-1842056. This work used
computational resources supported by the University of Tennessee and Oak Ridge
National Laboratory Joint Institute for Computational Sciences.
\end{acknowledgements}	
\appendix
	
\section{Results for the dynamical correlation functions\label{effectiveCorrelations}}

The interpretation of RIXS spectra computed with the Kramers-Heisenberg formalism can be difficult. 
To simplify matters, the full RIXS intensity can be expanded in powers of $J/\Gamma$ 
using the ultra-short core-hole lifetime (UCL) approximation. This procedure
expresses the RIXS intensity as a series of increasingly complicated multi-particle correlation functions, 
which can then be further subdivided into correlation functions of the NSC and SC channels. 
The detailed procedure to be followed has been reported in several prior studies ~\cite{UCL_expansion1, UCL_expansion2, PhysRevB.77.134428, PhysRevX.6.021020}. 
Here, we evaluate some effective correlation functions motivated by Eqs. (B1) and (B2) of 
Ref.~\onlinecite{PhysRevX.6.021020}. In many cases, these simplified
correlation functions give an accurate description of the RIXS intensity. 

\begin{figure}[t]
	\centering
	\includegraphics[width=1.01\linewidth]{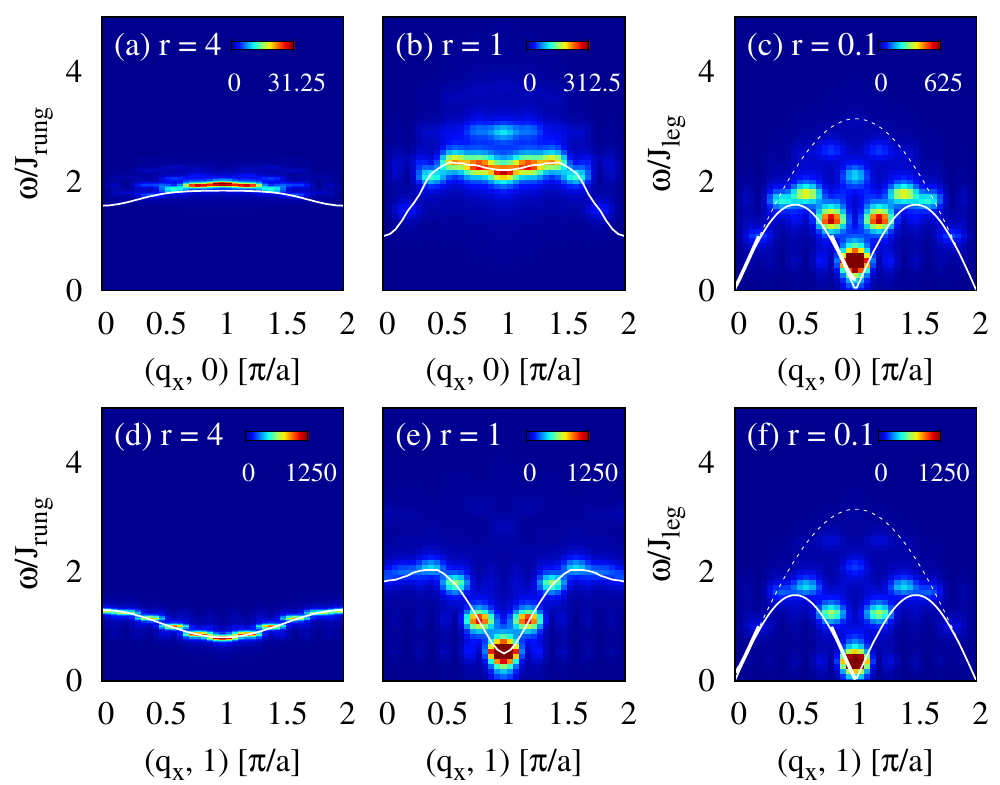}
	\vskip -0.250 cm        
	\caption{The dynamical spin correlation function  $S(\textbf{q},\omega)$ evaluated for 
        the undoped ladder with $r=\frac{J_\text{rung}}{J_\text{leg}}=\{4, 1, 0.1\}$ rung couplings, using ED and a 2$\times$10 cluster. 
        Panels (a)-(c) and (d)-(f) show the spectra 
        along the ${\bf q} = (q_x, 0)$ and ${\bf q} = (q_x, \pi/a)$ directions, respectively. The overall intensity 
        has been rescaled by a factor $1/\Gamma^2$, which corresponds to the prefactor relating $S(\textbf{q},\omega)$ 
        to the RIXS intensity. These plots capture all of the features of the spectra presented in Fig.~\ref{fig:rixsundopedS1}.}
	\label{fig:SqwS1}
\end{figure}

The spectral weight of the NSC channel is dominated by the first-order term in the UCL expansion, 
which is equivalent to the dynamical spin structure factor
\begin{equation}  \label{eq:Sqw}
S(\textbf{q},\omega) = \frac{1}{L} \sum_{f} \Big\vert \langle f|\sum_{i, \tau} 
e^{i\textbf{q}\cdot\textbf{R}_{i,\tau}} S_{i,\tau}^\alpha |g\rangle\Big\vert^2 \delta(E_f-E_g+\omega).
\end{equation}
Here, $S_i^\alpha$ ($\alpha =\{\pm,z\}$) is a component of the spin operator at site $(i,\tau)$.
The $S(\textbf{q}, \omega)$ responses for an undoped ladder along the $(q_x,0)$ and $(q_x, \pi/a)$ directions are plotted in 
Figs.~\ref{fig:SqwS1}(a)-(c) and ~\ref{fig:SqwS1}(c)-(f), 
respectively. These results compare well with the
RIXS intensity computed within the Kramers-Heisenberg formalism shown in
Figs.~\ref{fig:rixsundopedS1} for all values of the rung coupling.

To account for the magnetic excitations of the SC channel, the second-order term in the UCL expansion 
of the Kramers-Heisenberg formula is needed [see Eq.~(B2) of Ref.~\onlinecite{PhysRevX.6.021020}]. In the undoped case, 
the first-order term only contributes to the elastic line in this channel and a double spin-flip process 
appearing at second-order generates magnetic excitations. The RIXS spectra in the SC channel of the undoped 
ladders is hence dominated by the dynamical spin-exchange structure 
factor~\cite{PhysRevX.6.021020, PhysRevB.77.134428, PhysRevLett.106.157205} 
\begin{equation}\label{eq:S_exch}
\begin{split}
S^\text{exch}(\textbf{q},\omega) = \frac{1}{L} \sum_{f} \Big\vert \langle f| &\sum_{i, \tau} 
e^{i\textbf{q}\cdot\textbf{R}_{i, \tau}}  O_{i,\tau}^\text{exch} |g\rangle \Big\vert^2 \\& \times \delta(E_f-E_g+\omega).
\end{split}
\end{equation}
Here, $O_{i,\tau}^\text{exch} = \textbf{S}_{i,\tau}\cdot\left[J_\text{leg}\left(\textbf{S}_{i+1,\tau}+\textbf{S}_{i-1,\tau}\right)
+J_\text{rung}\textbf{S}_{i,\bar{\tau}}\right]/2$ evaluates the spin exchange between nearest-neighbor sites of the ladder, 
as sketched in Fig.~\ref{fig:schematicladder}(c) and \ref{fig:schematicladder}(d). 

\begin{figure}[t]
	\centering
	\includegraphics[width=\linewidth]{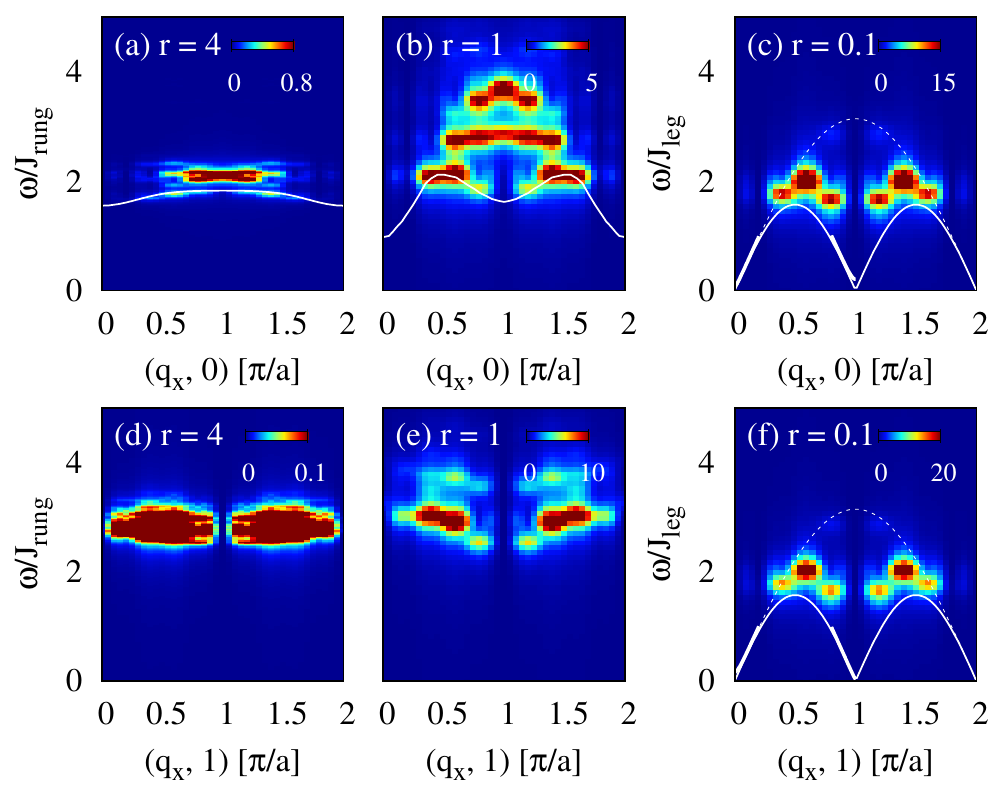}
	\caption{The dynamical spin exchange correlation function $S^\text{exch}(\textbf{q},\omega)$ 
        evaluated for the undoped ladder with $r=\frac{J_\text{rung}}{J_\text{leg}}=\{4, 1,0.1\}$ rung couplings, using ED and a 2$\times$10 cluster. 
 Panels (a)-(c) and (d)-(f) show the
spectra along the ${\bf q} = (q_x, 0)$ and ${\bf q} =  (q_x, \pi/a)$ directions, respectively. The overall intensity 
was rescaled by a factor of $1/\Gamma^4$, which corresponds to the prefactor 
relating $S^\text{exch}(\textbf{q},\omega)$ to the RIXS intensity. These
plots capture many of the features of the spectra Fig.~\ref{fig:rixsundopedS0}.}
	\label{fig:S_exchS0}
\end{figure}

The results for $S^\text{exch}(\textbf{q}, \omega)$ in the undoped ladder along the $(q_x, 0)$ and $(q_x,\pi/a)$ directions are 
shown in Figs.~\ref{fig:S_exchS0}(a)-(c) and ~\ref{fig:S_exchS0}(d)-(f),
respectively. These results compare well with the RIXS intensities shown in
Fig.~\ref{fig:rixsundopedS0} evaluted using the full Kramers-Heisenberg formalism.
$S^\text{exch}(\textbf{q}, \omega)$ captures the correct excitations for 
weak-rung couplings, but the intensities are overpredicted for strong-rung
couplings, the regime where the UCL approximation is expected to fail (large $J/\Gamma$).  

\begin{figure}
\centering
\includegraphics[width=\linewidth]{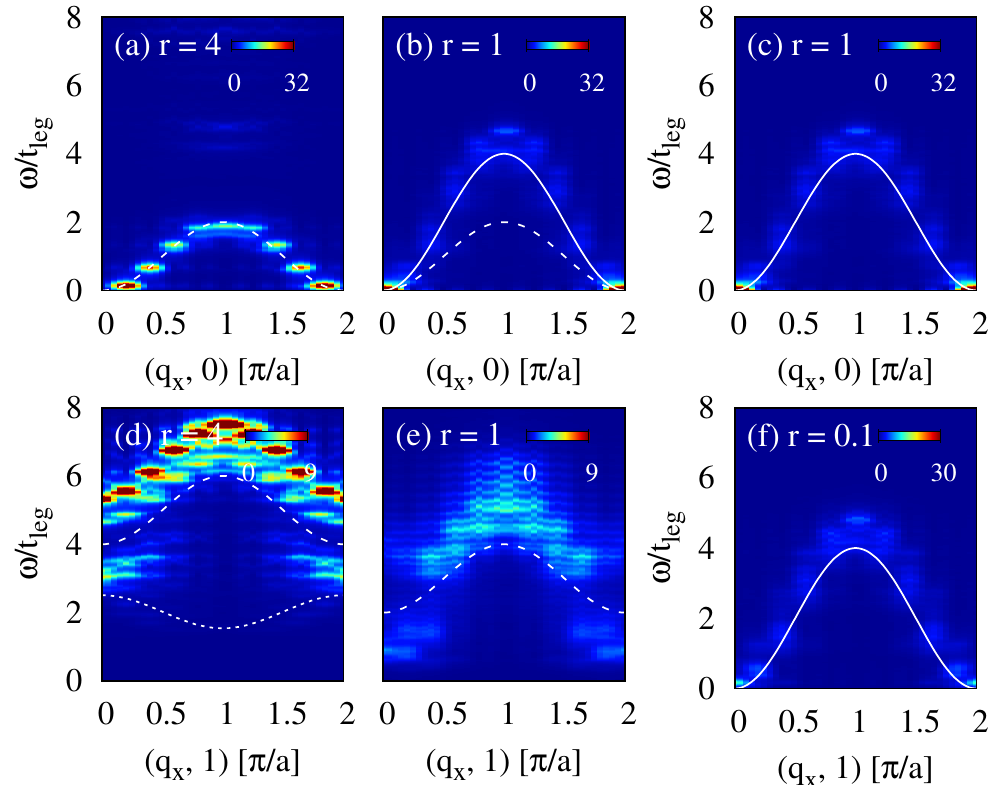}
\caption{The modified dynamical charge structure factor $\tilde{N}(\textbf{q},\omega)$ 
evaluated for the doped ladder with $r=\frac{J_\text{rung}}{J_\text{leg}}=\{4, 1, 0.1\}$ rung
couplings, using ED, a 2$\times$10 cluster, and $n=0.9$. Panels (a)-(c) and (d)-(f) show the spectra along the 
${\bf q} = (q_x, 0)$ and ${\bf q} = (q_x, \pi/a)$ directions, respectively. The overall intensity has been 
rescaled by a factor $1/\Gamma^2$, corresponding to the prefactor relating 
$\tilde{N}(\textbf{q},\omega)$ to the RIXS intensity. These plots capture many 
of the features of the spectra presented in Fig.~\ref{fig:rixsdopedS0}.} \label{fig:nqwdoped}
\end{figure}

The SC channel for the doped ladders is dominated by charge excitations.  
In this case, one must also go to second order and 
the RIXS intensity is well approximated by a modified dynamical charge structure 
factor~\cite{PhysRevX.6.021020} 
\begin{equation}\label{eq:nNqw}
\begin{split}
\tilde{N}(q,\omega) = \frac{1}{L}\Big( \sum_{f} \Big\vert \langle f| \sum_{i,\tau} & e^{i\textbf{q}\cdot\textbf{R}_{i,\tau}} O_{i,\tau}^\text{1} |g\rangle\Big\vert^2 \\ + \frac{1}{\Gamma^2}\Big\vert \langle f| \sum_{i,\tau} & e^{i\textbf{q}\cdot\textbf{R}_{i,\tau}} O_{i,\tau}^\text{2} |g\rangle\Big\vert^2\Big)
 \\ &\times\delta(E_f-E_g-\omega).
\end{split}
\end{equation}
Here, $O_{i,\tau}^\text{1} = \sum_{\sigma}c_{i,\tau,\sigma}^\dagger
c^{\phantom\dagger}_{i,\tau,\sigma}$,  $O_{i,\tau}^\text{2} = \sum_{\sigma}
[t_\text{leg}(c_{i+1,\tau,\sigma}^\dagger+ c_{i-1,\tau,\sigma}^\dagger)+
t_\text{rung}c_{i,\bar{\tau},\sigma}^\dagger]c^{\phantom\dagger}_{i,\tau,\sigma}/2$, and
$c_{i,\tau,\sigma}$ annihilates a spin $\sigma$ hole at 
site $(i,\tau)$, subject to the constraint of no double occupancy. Results for $\tilde{N}(\textbf{q}, \omega)$ along the $(q_x, 0)$ and
$(q_x,\pi/a)$ directions are shown in panels Fig.~\ref{fig:nqwdoped}(a)-(c) and 
Fig.~\ref{fig:nqwdoped}(d)-(f), respectively.  
In the doped case, the first-order term is non-zero but the majority of the intensity is set by the second-order term.  The $\tilde{N}({\bf q},\omega)$ results compare reasonably well to the 
RIXS spectra shown in Fig.~\ref{fig:rixsdopedS0} evaluated using the full Kramers-Heisenberg formalism.

\section{X-ray absorption\label{XAScalculation}}	
	\vskip -4mm
\begin{figure}[!htp]    
	\centering
	\hspace{-0.90cm}
	\begin{minipage}{0.49\linewidth}     
		\includegraphics[scale=0.80]{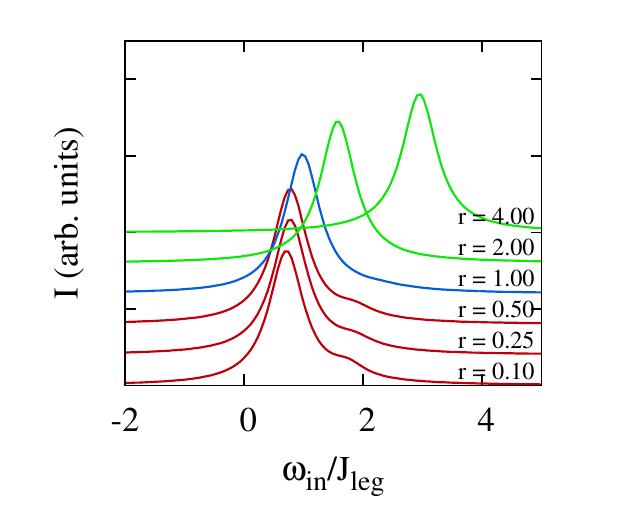}
	\end{minipage}
	\begin{minipage}{0.49\linewidth} \hspace{-0.90cm}
		\includegraphics[scale=0.80]{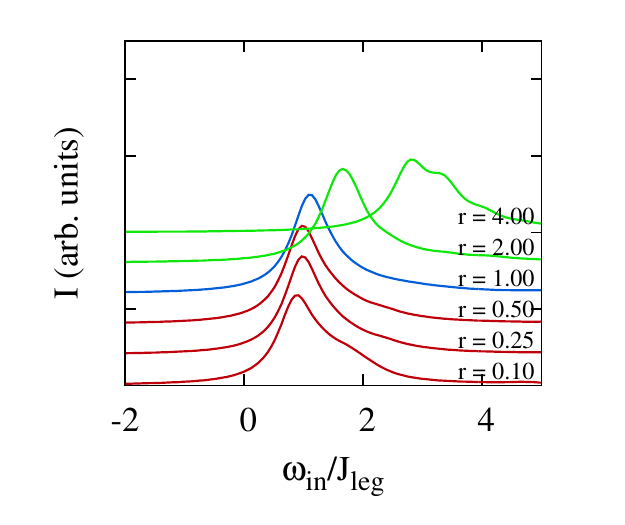}
	\end{minipage}    
	\vskip -4mm
	\caption{XAS spectra for the undoped~(left panel; $n=1.0$) and doped~(right panel; $n=0.9$) ladders  
		for various values of $r=\frac{J_\text{rung}}{J_\text{leg}}$, using ED and a 2$\times$10 cluster. Increasing the rung coupling factor $r$, 
		the peak position of the XAS shifts to higher incident photon energies. }
	\label{fig:xas}
\end{figure}

For all the RIXS figures discussed in the main text, 
we tuned the incident photon energy to the peak position observed in the x-ray absorption (XAS) spectra 
given by  
\begin{equation}
I(\omega_\mathrm{in}) = \sum_n \big|\langle n| D_{{\bf k} = 0} |g  \rangle \big|^2\delta(E_n  - E_g - \omega_\mathrm{in}).
\end{equation}

Figure \ref{fig:xas} shows the XAS spectra for the undoped and doped cases as a function of the rung 
coupling $r=J_\text{rung}/J_\text{leg}$. 
For the undoped two-leg ladder, the resonance peak in the XAS spectra shifts to larger values of 
$\omega_\text{in}$ with increasing rung coupling. In the strong-rung case ($r>1$), the ladder acts as a collection of rung-dimers and the shift in the XAS peak reflects the increased cost of breaking the dimer singlets. 
For the doped two-leg ladder, the overall intensity of the XAS also decreases with increasing $r$ while new peaks appear on the high-energy side of the resonance.

\bibliography{ladder}	

\end{document}